\documentclass[useAMS,usenatbib]{mn2e}

\usepackage{graphicx}
\usepackage{remreset} 
\usepackage{multicol}
\usepackage{amsmath}
\usepackage{float}
\usepackage{hyperref}

\usepackage{fixltx2e}

\def\apj{ApJ}
\def\mnras{MNRAS}
\def\aap{A\&A}
\def\apjs{ApJS}
\def\apjl{ApJL}
\def\aj{AJ}
\def\prd{PhRvD}
\def\pasa{PASA}

\def\nat{Nature}
\def\araa{ARA\&A}

\title[The Local Hole]{The Local Hole revealed by galaxy counts and redshifts}
\author[J.R. Whitbourn, T. Shanks]{J.R. Whitbourn$^{1}$\thanks{E-mail:
(JRW) jrwhitbourn@gmail.com} and T. Shanks$^{1}$\thanks{E-mail: (TS)
tom.shanks@durham.ac.uk} \\
$^{1}$Extragalactic \& Cosmology Group, Department of Physics, Durham
University, South Road DH1 3LE}

\begin{document}
\date{Accepted 2013 October 18; Received 2013 October 7; in original form 2013 July 16}

\pagerange{\pageref{firstpage}--\pageref{lastpage}} \pubyear{2013}
\maketitle

\label{firstpage}


\begin{abstract}

The redshifts of $\approx250000$ galaxies are used to study the Local
Hole and its associated peculiar velocities. The sample, compiled from
6dF Galaxy Redshift Survey (6dFGS) and Sloan Digital Sky Survey (SDSS), provides
wide sky coverage to a depth of $\approx300$h$^{-1}$Mpc. We have therefore
examined $K$ and $r$ limited galaxy redshift distributions and number counts to
map the local density field. Comparing observed galaxy $n(z)$ distributions to
homogeneous models in three large regions of the high latitude sky, we find
evidence for under-densities ranging from $\approx$4-40\% in these regions to
depths of $\approx$150h$^{-1}$Mpc with the deepest under-density being over the
Southern Galactic cap. Using the Galaxy and Mass Assembly (GAMA) survey we then
establish the normalisation of galaxy counts at fainter magnitudes and thus
confirm that the underdensity over all three fields at $K<12.5$ is
$\approx15\pm3$\%. Finally, we further use redshift catalogues to map
sky-averaged peculiar velocities over the same areas using the average
redshift - magnitude, $\overline{z}(m)$, technique of \citet{soneira_1979}.
After accounting for the direct  effect of  large-scale structure on
$\overline{z}(m)$ we can then search for peculiar velocities. Taking all three
regions into consideration the data reject at the $\approx4\sigma$ level the
idea that we have recovered the CMB rest frame in the volume probed. We
therefore conclude that there is some consistent evidence from both counts and
Hubble diagrams for a `Local Hole' with a $\approx150$h$^{-1}$Mpc under-density
that deeper counts and redshifts in the Northern Galactic cap suggest may extend
to $\approx300$h$^{-1}$Mpc.

\end{abstract}



\section{Introduction}
\label{sec:intro}

The Cosmological Principle is a fundamental assumption of cosmology that
leads us to describe our universe as statistically homogeneous and isotropic,
which uniquely gives the Friedmann-Lemaitre-Robertson-Walker (FLRW) solutions to
Einstein's field equations. These metrics are apparently successful,
encompassing many current observations of the Universe over huge scales in
space, time and energy.

However, at least locally, the validity of the Cosmological Principle is less
obvious. Deep redshift surveys such as SDSS \citep{york_2000} and 2dFGRS
\citep{colless_2001} have revealed a web-like structure to the galaxy
distribution with extensive and ongoing clustering at knots and
junctions. Indeed, recent redshift surveys have found this Large Scale
Structure (LSS) persisting up to at least scales of $300 h^{-1} Mpc$
\citep{gott_2005,murphy_2011}. The results are in concordance with
$\Lambda CDM$ N-body simulations with the galaxies displaying the
expected hierarchical structure from individual galaxies to galaxy
clusters to superclusters \citep{park_2012,watson_2013}. The visible structures are
parsed by large coherent regions of under-density known as voids, which
can be of $\mathcal{O}(50Mpc)$. Compared to galaxy clusters, voids were
a relatively recent discovery in cosmography as they required large
redshift surveys to easily separate galaxies in the same line of sight
by redshift. These regions seem to be approximately spherical and underdense in all
types of matter \citep{peebles_2010,rood_1988}.

The question of the local galaxy density has received renewed attention due to the 
challenges represented by the recent measurements of a $\Lambda$-like
accelerated expansion of the universe \citep{schmidt_1998,perlmutter_1999}. There
is the possibility that the role of $\Lambda$ in producing the dimming
of the $m-z$ relationship for SN1a could instead be due to the
acceleration induced by a large local under-density.
Recently it has been shown that $\mathcal{O}(Gpc)$ local hole models can
accurately mimic $\Lambda$ whilst accounting for independent scale factor
measurements \citep{february_2010}. However, it remains unclear as to
whether these models can equally well simultaneously account for other
cosmological datasets\footnote{Baryon Acoustic Oscillations (BAO), H(z), kSZ,
Lithium Abundance, CMB fluctuations and Cosmic Shear} - see \cite{biswas_2010,
moss_2010} and also  \cite{regis_2010,nadathur_2011}.

\subsection{Scale of Homogeneity}

Results disagree  as to whether recent redshift surveys have
approached the depths required to describe the universe as statistically
homogeneous. Studies of the fractal dimension of the galaxy distribution
typically report a homogeneity scale of $\approx70$h$^{-1}$Mpc
\citep{sarkar_2009,scrimgeour_2012,hogg_2005}. However, other studies instead
find the presence of LSS beyond these scales and indeed persisting to the
relevant survey depths \citep{labini_2011,celerier_2005}. 

Efforts to use the number or flux dipole in a similar manner to the peculiar
velocity dipole have been in concordance with the $\Lambda CDM$ standard model
\citep{bilicki_2011,blake_2002}. \citet{gibelyou_2012} report that the NVSS
number dipole is unexpectedly large, however they attribute this to potential
systematic errors.

Studies of the structure of our local peculiar velocity field have used
the scale at which the bulk peculiar velocity is that of the CMB dipole
as a proxy for the scale of homogeneity. Some authors have reported a
relatively local origin within $\approx60$h$^{-1}$Mpc for the dipole
\citep{erdogdu_2006}. However, other recent studies have suggested that there
are bulk flows at much larger scales
\citep{abate_2012,watkins_2009,feldman_2010,colin_2011}. These results are in
contrast with a series of papers, \citep{nusser_2011,nusser_2012}, where a
method similar to one used here is pioneered and bulk flows consistent with
$\Lambda CDM$ were found. 

Furthermore, attempts to infer the bulk velocity field with respect to the CMB
have typically returned values incompatible with homogeneity
\citep{kashlinsky_2008,lavaux_2012}. These results are however disputed by some
authors \citep{keisler_2009,osborne_2011}.

\subsection{Number Counts}

By counting the number of galaxies as a function of magnitude and
redshift, strong constraints can be imposed on galaxy evolution, galaxy
distribution and cosmology. The existence of LSS in the form of superstructures
such as filaments can be readily detected in these counts
\citep{frith_2003}. 

In the  standard model with $\Lambda$, number counts for $z < 1$ are
well described by simple Pure Luminosity Evolution (PLE) models where
galaxies form at  high redshift  and evolve according to their galaxy 
star-formation rate, with  e-folding times assumed to be
$\tau=1-2.5$Gyr for redder types and $\tau=9$Gyr for bluer types. These
PLE models are successful across a wide range of passbands and to
considerable redshift depth
\citep{shanks_1984,metcalfe_2001,metcalfe_2006,hill_2010}

However, the above PLE models cannot simultaneously account for bright
and faint magnitude counts \citep{liske_2003,metcalfe_2001}.
Specifically, the counts in the range $10<B<17$mag are significantly
steeper than expected from a non-evolving model. Indeed the counts at
fainter magnitudes are less steep relative to such a model. As long as
the PLE model counts were normalised at $B\approx18$mag the PLE models
then fit in the range  $18<B<28.5$ \citep{metcalfe_2001} but attempts to
fit at $B<17$ inevitably overshoot beyond $B>17$ and it seemed puzzling
that the evolution rate should increase at lower redshift. It was
therefore suggested that the  steepness of the bright counts may be
caused by a local under-density \citep{shanks_1984}. Luminosity functions (LF) 
measured in redshift surveys are reasonably consistent in form but there exists
considerable variation in $\phi^{*}$ \citep{liske_2003,cross_2001}. This
uncertainty is in part due to the failure of  non-evolving (or simple PLE)
models to fit  bright and faint counts simultaneously and is known as the
normalisation problem. There is supporting evidence for a faint count 
normalisation from several previous studies \citep{driver_1995,
glazebrook_1995},  complemented by results from  the latest and deepest number
counts \citep{keenan_2010, barro_2009} and luminosity functions
\citep{keenan_2012,keenan_2013}.

A further  argument against the steep bright counts being caused by
$z<0.1$ galaxy evolution is that the steepness is observed across the
NIR and optical bands (B,R,I,H,K) \citep{metcalfe_2001,metcalfe_2006}.
In models where SFR dominates the evolution, we should expect the bluer
bands to be more affected than the redder bands and this effect is seen
at fainter magnitudes but not at brighter magnitudes.

Using early (partial) 2MASS data releases, Frith embarked on a series
of analyses at bright NIR magnitudes to investigate the strong local LSS
hypothesis. \citet{frith_2003} observed evidence for the reality of the
proposed local under-density with the underdensities in the 2DFGRS
redshift distribution accounting well for the underdense 2MASS number
counts - see also \citealt{busswell_2004}. The galaxy distribution was found to
be patchy with large regions of under- and over-density. Across the whole sky a
coherent $\approx15-20$\% under-density, a local hole, on the scale of
$\mathcal{O}(300Mpc)$ was consistent with these data.

\citet{frith_2006a} also found further evidence that the faint
normalisation is correct in the H band. Using a set of 2MASS mocks, 
the full sky under-density was found to represent a 2.5$\sigma$ fluctuation
for a $\Lambda$CDM model.

In this paper we attempt to extend the \citet{frith_2005b} analysis of the local hole
hypothesis. We first check out the connection between  $n(z)$ and $n(m)$
in substantially bigger areas than available to Frith et al. We also  
test whether there is an under-density in the mass as well as the galaxy 
counts by estimating a velocity field using the \citet{metcalfe_2001}
luminosity function and the $\overline{z}(m)$ Hubble diagram technique of
\citet{soneira_1979} which we outline below.

\begin{figure*}
\begin{center}
 \includegraphics[width=11.cm]{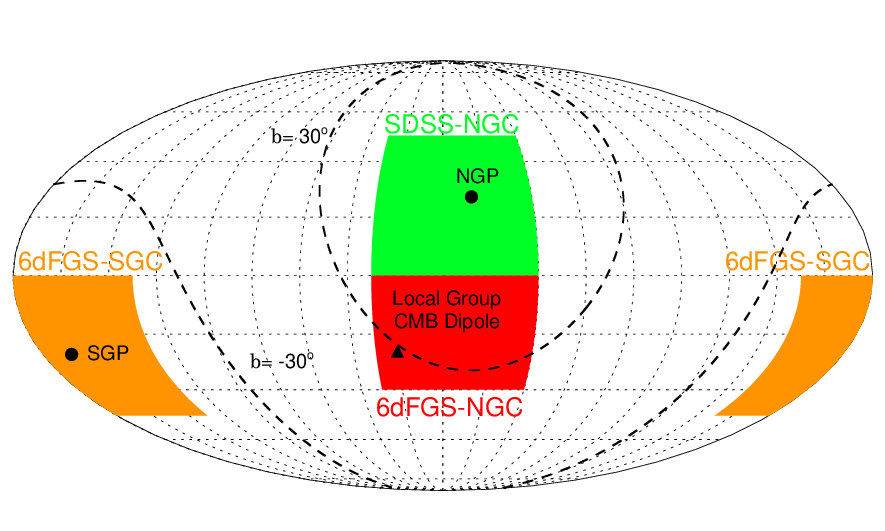}
 \caption{A Mollweide projection of the fields used in this study using the
celestial coordinate system. The 6dFGS-NGC field is represented by the filled area in red, the 6dFGS-SGC
field is represented by the filled area in orange and the SDSS-NGC is represented by the filled
area in green. Also shown are the Northern and Southern Galactic poles, the
Local Group CMB dipole pointing and lines of $b=30^\circ$ and $b=-30^\circ$ galactic
latitude.}
 \label{fig:skymap}
\end{center}
\end{figure*}

\section{Techniques}

\subsection{Number-magnitude and number-redshift distributions}
\label{subsec:tech-nmnz}

We will first compare the number-redshift and number-magnitude distributions
with those that assume homogeneous models. We assume simple LFs as described by
\citet{metcalfe_2001} and so predict the differential number redshift relation
$n(z)$ using

\begin{equation}
 n(z)dz=4\pi r(z)^2\frac{dr}{dz}dz\int_{-\infty}^{M(m_{lim},z)}{\Phi(M)}dM,
 \label{eq:redshiftdistimplement}
\end{equation}

\noindent where $m_{lim}$ is the survey magnitude limit, $r(z)$ is the
comoving radial coordinate, $\Phi(M)$ is the differential \citet{schechter_1976} luminosity function in comoving units with characteristic
absolute magnitude and density, $M^*(z)$ and $\phi^*(z)$ and slope
$\alpha$. Our models for the redshift dependence of $M^*(z)$ include K
plus E corrections from Bruzual \& Charlot (2003) models with $\phi^*$
and  $\alpha$ held constant for individual galaxy types for the
homogeneous models. We shall generally normalise the homogeneous $n(z)$
model to exceed the observed $n(z)$ by the ratio of homogeneous model
counts to the observed $n(m)$. We shall then simply divide the observed
$n(z)$ by the homogeneous model $n(z)$ to determine how the galaxy
density $\phi^*(z)$ varies with redshift.

The homogeneous number-magnitude relation is then similarly calculated as,

\begin{equation}
 n(m)\Delta m = \int_0^\infty{4\pi r(z)^2\frac{dr}{dz}dz}\int_{M(m_b,z)}^{M(m_f,z)}{\Phi(M)}dM,
 \label{eq:magnitudedistimplement}
\end{equation}

\noindent where $m=(\frac{m_b+m_f}{2})$, $\Delta m= m_f-m_b$. 
We can then also input $\phi^*(z)$ from $n(z)$ into the $n(m)$ model to check for consistency 
between any under- or over-densities found in $n(z)$ and $n(m)$.

\subsection{Hubble Diagrams from galaxy redshift surveys}

Hubble's law relates cosmological redshifts to  distance. Usually the
distances come from standard candles or rods for individual galaxies.
But here we aim to use the galaxy luminosity function as the standard
candle for magnitude limited samples of galaxies using the average
redshift  as a function of magnitude, $\overline{z}(m)$, following
\citet{soneira_1979}. In essence the method assumes a universal LF which
is an approximation, ignoring environmental effects etc. But the bigger
the volumes averaged the more this assumption will apply and the LF can
then be used as a statistical standard candle. 

\citet{soneira_1979} working at small redshifts, assumed a Euclidean cosmology
and  the redshift-distance relation, $z=br^{p} + y$, where the
peculiar velocity $y$ distribution is described by $Q(y)$, and derived

\begin{equation}
 \overline{z}(m) \propto 10^{0.2pm}.
\label{eq:soneiraexpectagain}
\end{equation}

\noindent Clearly for a linear Hubble law, $p=1$ and the aim of Soneira's 
analysis was to determine $p$. Here we use the same technique out to higher
redshift where the potential effects of cosmology, K correction and evolution
cannot be ignored. We can describe $\overline{z}(m)$ in complete generality
using the volume element $dV/dr$, the differential LF $\Phi(M)$, the peculiar
velocity distribution $Q(y)$ and K plus E corrections,

\begin{equation*}
\overline{z}(m) = \frac { \int\limits^{\infty}_{\rm -\infty} dy \int\limits^{\infty}_{\rm 0}
z(r,y) Q(y) \Phi(m-5log d_L-25-KE(z)) \frac{dV}{dr} dr}  {\int\limits^{\infty}_{\rm -\infty}
dy \int\limits^{\infty}_{\rm 0} Q(y) \Phi(m-5log d_L-25-KE(z)) \frac{dV}{dr} dr}.
\label{eq:soneriageneral0}
\end{equation*}

\noindent We initially only make the simplest set of assumptions about
$Q(y)$, that it is normalised to one and with a mean of zero, i.e: non-streaming,

\begin{equation}
 \int^{\infty}_{\rm -\infty} Q(y)  dy = 1, \ \  \int^{\infty}_{\rm -\infty} y Q(y)  dy = 0 \label{eq:soneriacond_1}.
\end{equation}

\noindent In the case of velocity flows we have more complicated forms of
$Q(y)$. The simplest such case is a bulk flow where all galaxies are moving
coherently,

\begin{equation}
\int^{\infty}_{\rm -\infty} y Q(y)  dy = \frac{v_{\rm flow}}{c}.  \\
\label{eq:soneirapecvelsimp}
\end{equation}

\noindent The implication for $\overline{z}(m)$ is that,

\begin{equation}
\overline{z}(m) = \overline{z}_{hubble}(m) + \frac{v_{\rm flow}}{c}.
\label{eq:soneirapecvelsimpzbar}
\end{equation}

\noindent Therefore $\overline{z}(m)$ is dependent on galaxy streaming velocity. 
                                 
$\overline{z}(m)$ is calculated in magnitude bins. We have chosen to use
both $\delta m=0.5$ and $\delta m=0.1$. The larger $\delta m=0.5$ binning
is preferred because these  have slightly smaller  errors and reduced covariance
between bins. However, we have also presented results for $\overline{z}(m)$ with
the smaller magnitude binning size of $\delta m = 0.1$  to investigate the
sensitivity of $\overline{z}(m)$ to individual elements of LSS, which
the larger binning suppresses.

\section{Modelling}
\label{sec:kecorrections}

We now need to model $n(z)$,  $n(m)$ and $\overline{z}(m)$ first in the
homogeneous case so below we present details of the galaxy evolution
models and the luminosity function parameters. 

\subsection{Galaxy Evolution Models}

A galaxy's apparent magnitude is dependent on both evolution and SED, hence
modelling $\overline{z}(m)$ requires us to account for the $k(z)$ and $e(z)$
effects. The K plus E corrections used in this paper are calculated using the
stellar synthesis models set out in \citet{bruzual_2003}. We have used an $x=3$
IMF for early types to mimic the PLE galaxy models set out by
\citet{metcalfe_2001,metcalfe_2006}.

In this paper we will usually present results in the NIR and at low redshift,
where the $e(z)$ and $k(z)$ corrections are relatively small and can be
reasonably well determined. This is because the NIR is dominated by old stars
and hence is insensitive to different star formation histories.
\citep{bruzual_2003,cole_2001}. We have verified this by
experimenting with alternative forms for the $k(z)$ and $e(z)$ correction and
found that the results are not sensitive to the exact form used.

\subsection{Luminosity Functions}

Our basic LF will be taken from \citet{metcalfe_2001}. This is a type
dependent LF that is inferred from the optical and translated into the NIR using
the mean colours - see Table 1. Modelling of the number counts, redshift distributions and
$\overline{z}(m)$ using this LF has been done using the full number count
program described by \citet{metcalfe_1996}.

\begin{table}
\begin{center}
 \begin{tabular}{ccccc}

\hline\hline
Type & $\phi \ (h^{3} Mpc^{-3})$ & $\alpha$ & $M^{*}_{\rm R} - 5\log(h)$ & R-K \\
\hline\hline

E/S0 & $7.416 \cdot 10^{-3}$ & -0.7 & -20.93 &  2.48 \\

Sab & $3.704 \cdot 10^{-3}$ & -0.7 & -20.75 &  2.52 \\

Sbc & $4.960 \cdot 10^{-3}$ & -1.1 & -20.87 &  2.45 \\

Scd & $2.184 \cdot 10^{-3}$ & -1.5 & -20.70 &  2.13 \\

Sdm & $1.088 \cdot 10^{-3}$ & -1.5 & -20.62 &  1.58 \\

\hline\hline
\end{tabular}
\caption{Parameters for the zero redshift luminosity function as assumed here
\citep{metcalfe_2001, metcalfe_2006}. We will use a $\Lambda$CDM cosmology with
$\Omega_{\Lambda}$=0.7, $\Omega_{m}$=0.3 and $h=0.7$.}
\end{center}
\label{tb:shankslf}
\end{table}

\subsection{Radial Inhomogeneity - LSS Correction}
\label{subsect:fongeffectjust}

The derivation of $\overline{z}(m)$ shown earlier assumes radial homogeneity,
$\overline{z}(\overline{r})=\overline{z}(r)$ which  leads to a sensitivity to
over/under-densities, as was indeed originally noted by
\citet{soneira_1979}. For example, the presence of a local hole would be
expected to cause a boost to $\overline{z}(m)$ at bright magnitudes  (small distances), 
even with no induced peculiar motion. This is because at a bright
apparent magnitude, $m$, the ratio of galaxies outside the hole (with high
$z$) and galaxies inside the hole (at low $z$) would be expected to increase
with hole density contrast and scale. The inverse would be expected in the
presence of a local over-density.

We can model this effect by varying the normalisation, $\phi^*$ of the LF we
use. To do this we will include radial density profiles derived from our $n(z)$
distributions. Rather than allowing this measure to extend to the survey limits
where the effect of redshift incompleteness and survey systematics become more
prominent, we set a scale, $z_{\rm global}$ where we transition to the
expected homogeneous value. We use values of $z_{\rm global}$=0.15 and $z_{\rm
global}$=0.25 for the K and r bands respectively.

\begin{equation}
\phi^*(z) = %
\begin{cases}
\frac {n(z)_{\rm obs}}{n(z)_{\rm model}}\phi^*_{\rm global} & \text{if } z \leq
z_{\rm global}  \\
\phi^*_{\rm global} & \text{if } z > z_{\rm global}
\end{cases}
\label{eq:fongphipiecewise}
\end{equation}

We are assuming the density variations in the $n(z)$ are real and using this
to correct the $\overline{z}(m)$ model prediction for the effect of  large-scale
structure before looking for residuals that can be interpreted as peculiar velocities,
$v_{pec}$. We shall also use the same technique to correct our homogeneous model $n(m)$
prediction for the effect of large-scale structure to make  consistency checks
between $n(m)$ and $n(z)$.

In a following paper, \citet{shanks_2013} will use simple
maximum likelihood estimates of the luminosity function also to estimate
$\phi^*(z)$ simultaneously. We find that the \citet{metcalfe_2001} LF used here
is in good agreement with these ML estimates. The $\phi^*(z)$ density runs
with redshift also agree with those reported below.

\subsection{Error calculation}
\label{subsec:errors}

As a first approximation it is possible to assume Poisson errors for
the number counts and standard errors for $\overline{z}(m)$. This though
is unrealistic for real galaxy distributions since galaxies cluster. 
To account for this we have therefore calculated jack-knife errors. These were
calculated using $10^\circ\times10^\circ$ sub-fields. For $N$ fields denoted
\textit{k}, the errors on a statistic $f$ as a function of the variable $x$ are,

\begin{equation}
 \sigma_{f}^{2}(x) = \frac{N-1}{N} \sum_{k}^{N} \Big( f_{k}(x) - \overline{f}(x) \Big)^{2},
\label{eq:errdefintion}
\end{equation}

\noindent where $f_{k}(x)$ is the average of the fields excluding field $k$. We
have experimented with both more survey specific sub-fields and alternative
methods such as field-to-field resampling and find approximately equivalent
results in these cases.

\section{Data -  Surveys}
\label{sec:data}

In this section a compilation is given of the key characteristics of the
imaging and redshift surveys used throughout this work. We shall generally use
pseudo-total  magnitudes, usually estimated by integrating a fitted
analytic surface brightness profile to large radii - for details see
individual surveys below. We shall use magnitudes zeropointed in the
$Vega$ system throughout. This  is primarily for ease since  the $2MASS$
photometry is quoted in this system. Where necessary we have converted
from AB to Vega using the following offsets from \citet{hill_2010} and
\citet{blanton2007},

\begin{align}
 K_{vega} &= K_{AB} - 1.90, \\ \nonumber
 r_{vega} &= r_{AB} - 0.16.
\label{eq:vega_ab_conv}
\end{align}
 
The NIR is minimally affected by dust extinction but we have  applied extinction
corrections using the extinction maps of \citet{schlegel_1998}. We note that our results 
are insensitive to whether we apply the correction at all. This applies in $r$ as well 
as $K$ since the $r$ band data used below are restricted to  higher galactic latitudes

In terms of the redshift surveys,  we choose to work in the Local group rest-frame. All redshifts
have therefore been corrected to the Local group barycenter using
$(l_{LG},b_{LG})=(93^{\circ},-4^{\circ})$ and $v_{LG}=316 kms^{-1}$
\citep{karachentsev_1996},

\begin{align}
 cz_{LG} = cz_{\sun} + &v_{LG} \big[ \sin(b)\sin(b_{LG}) \\ \nonumber
 & + \cos(b)\cos(b_{LG})\cos(l-l_{LG}) \big].
 \label{eq:zconv_helio_lg}
\end{align}

\subsection{Imaging Surveys}

We next discuss the main characteristics of the imaging surveys used in this work. 
The details of the tests we have done on the magnitude scales, star-galaxy separation etc
are given in Appendix \ref{append:magaccuracy}.

\subsubsection{2MASS}

The Two Micron All-Sky Survey (2MASS, \citet{skrutskie_2006}) is a photometric
survey in the NIR (J,H,$K_{\rm s}$). The final eXtended
Source Catalogue (2MASS-XSC) comprises of 1,647,459 galaxies over approximately
the whole sky (99.998\% sky coverage), with a photometric calibration varying by
as little as 2-3\% \citep{jarrett_2003}. 2MASS is currently thought to be
magnitude complete to $K < 13.5$ \citep{bell_2003,chodorowski_2008}. 

The 2MASS-XSC data used in this paper comes from the `All-Sky Data Release' at
the IPAC server. Galaxies have been included according to the following quality
tags: `$\textit{cc\_flg}=0$', `$\textit{cc\_flg}=Z$' to avoid contamination or
confusion. The XSC catalogue consists solely of 2MASS objects
with e-score and g-score $< 1.4$ to ensure the object really is extended and
extragalactic.

It has been reported that the completeness and photometry of 2MASS-XSC galaxies
with angular diameter greater than 10$'$ may be affected by the limit on the
2MASS scan size \citet{jarrett_2003}. We have therefore applied a bright magnitude cut
of K$>$10 for $n(m)$, $n(z)$ and $\overline{z}(m)$.

For the 2MASS survey, we shall use a corrected form (see Appendix
\ref{append:magaccuracy}) of their extrapolated isophotal,
\textit{k\_m\_ext},  magnitude. This total type magnitude is based on an
integration over the radial surface brightness profile. The lower radial
boundary is defined by the isophotal $\mu$ = 20mag $arcsec^{-2}$ radius
and an upper boundary by four disk scale lengths unless that is greater than 
5 of the above minimum isophotal radii.

\subsubsection{GAMA}

The Galaxy And Mass Assembly (GAMA, \citet{driver_2009}) survey includes
galaxies selected from UKIDSS-LAS and SDSS photometric targeting. It
aims to create a catalogue of $\approx 350,000$ galaxies with
comprehensive photometry from the UV band to the radio. GAMA DR1  is
based on three 45deg$^2$ equatorial regions, chosen for their overlap
with SDSS(stripes 9-12) and UKIDSS-LAS data. It comprises self
consistent (ugrizJHK) imaging of 114,441 galaxies with 50,282 science
quality redshifts.  

As of GAMA DR1, only the Kron type K magnitude, \textit{K\_KRON} has
been provided, and therefore we use this magnitude type. Whilst the NIR
GAMA photometric data comes from UKIDSS, the final catalogue has been
re-reduced for a variety of reasons outlined by \citet{hill_2010}.  

The GAMA data used here comes from the DR1 release, GAMACoreDR1,
described by \citet{driver_2011} and archived at
(\path{http://www.gama-survey.org/database/YR1public.php}). We
have selected all galaxies in GAMA DR1, including those based on band
specific detections.

\subsubsection{SDSS}

The Sloan Digital Sky Survey (SDSS, \citet{york_2000}) covers $\approx 8500
deg^{2}$ of the Northern sky in the u,g,r,i,z bands. As of DR9 the survey
comprises 208,478,448 galaxies and is magnitude complete to $r_{petro}<$22.04.

For consistency, we have chosen to work with the same magnitude type for
both spectroscopic and photometric SDSS samples. We therefore use the
`cmodel' type magnitude as recommended by SDSS
\footnote{\path{https://www.sdss3.org/dr8/algorithms/magnitudes.php\#which
\_mags}}. This total type magnitude is estimated by determining de
Vaucouleurs or exponential profiles for each object in each band. The
likelihood of either profile is then determined and the linear
combination that best fits is then used to infer the total flux.
A photometric sample has been selected using the quality criteria developed by
\citet{yasuda_2001} for galaxy number counts. Namely, we reject saturated and
non-primary objects and require a photometric classification as a galaxy in at
least two of the $g,r,i$ bands.

\subsection{Redshift Surveys}
\label{sec:dataredshift}

Next we describe the main characteristics of the redshift surveys used in this
work. In Appendix \ref{append:magincompleteness} we discuss the tests we have
made on the magnitude dependent spectroscopic incompleteness of these surveys
and how such effects can be corrected in the redshift distributions, $n(z)$.

\subsubsection{6dFGS}

The Six Degree Field Galaxy (6dFGS, \citet{jones_2004}) is a redshift
survey over $\approx 17,000 deg^{2}$ i.e: most of the Southern sky,
excluding $\left| b \right| < 10$. The survey was based on pre-existing
overlapping survey photometry and was primarily selected in 2MASS K. The
full survey comprises a catalogue of 125,071 galaxies with reliable
redshifts. The survey has a median redshift of $z_{\rm median} =0.053$
\citep{jones_2009} to its nominal limit of $K\leq12.65$. We, however,
shall be conservative and impose a $K < 12.5$ magnitude cut to minimise
any completeness issues with the 6dFGS data. The 6dFGS data used in this
paper comes from the final DR3 release described in \citet{jones_2009}
and is archived at (\path{http://www-wfau.roe.ac.uk/6dFGS/}). Galaxies
have been included according to the following quality tags; $\textit{quality}
\geq 3$, $\textit{quality} \ne 6$.

It is historically relevant to note that the 6dFGS survey was started
before the final 2MASS photometry was released. Intermediate 2MASS
photometry at low galactic latitudes was relatively shallow and suffered
from poor spatial resolution. To work around this the 6dFGS team adopted
a pseudo-total magnitude for redshift targeting. Other researchers used
an alternative J-K inferred isophotal magnitude, hence referred to as a
Cole type \citep{cole_2001}. With this type of estimator the less noisy
J band is used to approach the true K band magnitude as $K_{cole} =
J_{ext} - (J_{iso} - K_{iso})$.  This type was indeed found to have
greater accuracy compared to the accurate photometry of
\citet{loveday_2000}. However, the final release of the 2MASS catalogue
provided the total estimator, \textit{k\_m\_ext} as described earlier.
The 6dFGS team recommend this magnitude for science use. However, it
remains the case that 6dFGS was targeted in a slightly different
magnitude and that previous work has been conducted in a variety of
magnitudes.

\subsubsection{SDSS - Spectroscopic Survey}

The spectroscopic sample was selected to a limit of $r_{petro} < 17.61$ finally
comprising 1,457,002 confirmed galaxy redshifts, with a median redshift, $z_{\rm
median} = 0.108$. The SDSS spectroscopic sample was targeted on the
basis of Petrosian magnitudes \citep{strauss_2002}. We however are working with
the cmodel type magnitude. To avoid selection and completeness effects we
therefore choose to work with the conservative magnitude limit
$r_{cmodel}<$17.2.

We have also created a K limited SDSS spectroscopic sample by matching with
2MASS. The SDSS astrometric error is of order $\mathcal{O}(0.1'')$
\citep{hill_2010,finlator_2000} we therefore set a $1''$ matching limit. For this K
limited SDSS sample we are in effect applying the
multi-band selection that $K<13.5$ and $r<17.61$. This additional
constraint does not bias the sample we select since even for a galaxy at the
2MASS limit it will require a relatively blue r-K colour of 4.11 to avoid
selection in the joint sample. Indeed, \citet{bell_2003} found that at most 1\%
of galaxies were affected in a similar joint SDSS-2MASS sample.

The SDSS data used in this paper come from the DR9 main sample described in
\citet{ahn_2012} and is archived at
(\path{http://skyserver.sdss3.org/CasJobs/}) In order to select a fair
and high quality sample of galaxies we have used the following selection
criteria; $\textit{class=`GALAXY'}$, $\textit{(zWarning = 0
OR ((zWarning} \& (4)) \ge 0)$, $legacy\_target1 \& (64 | 128 | 256)) \le 0$,
$\textit{mode=1}$ and $\textit{scienceprimary=1}$

\begin{table}
 \begin{center}
 \begin{tabular}{ccccc}

\hline\hline
Survey & $z_{\rm median}$ & Mag limit & Area ($deg^{2}$) \\
\hline\hline

6dFGS & 0.053 & $K_{\rm s}<12.5$ & 17000 \\

SDSS-MAIN & 0.108 & $r<17.61$ & 8500 \\

GAMA & 0.18 & $r<19.24$ & 150 \\

\hline

2MASS & - & $K_{\rm s}<13.5$ & $\sim$ Full Sky \\

SDSS-MAIN & - & $r<22.04$ & 8500 \\

\hline\hline
\end{tabular}
\caption{A summary of the properties of the redshift and imaging surveys
used; (6dFGS, \citet{jones_2004}), (SDSS,
\citet{york_2000}) (GAMA, \citet{driver_2009}) and (2MASS, \citet{jarrett_2003}).}
 \end{center}
 \label{tb:galsurveysredshift}
\end{table}

\subsection{Target Fields}
Three fields were chosen to cover most of  the northern and southern
galactic caps at high latitudes while maintaining the basic division
between the northern SDSS and southern 6dFGS redshift survey areas, as
shown in Fig. \ref{fig:skymap} and Table {\ref{tb:fields}}. The three
fields are termed  SDSS-NGC, 6dFGS-NGC and 6dFGS-SGC as shown in Fig.
\ref{fig:skymap}. These fields  contain various regions of interest. The
6dFGS-NGC contains the CMB Local group dipole pointing,  the direction
of  Great Attractor and the Shapley-8 supercluster. The 6dFGS-SGC region
contains the Perseus-Pisces supercluster, whilst the SDSS-NGC region
contains the Coma cluster.

\begin{table}
 \begin{center}
 \begin{tabular}{cccc}

\hline\hline
Field & RA (J2000) & DEC (J2000) & Area ($deg^{2}$) \\
\hline\hline

6dFGS-NGC & [150,220] & [-40,0] & 2578.03 \\

6dFGS-SGC & [0-50,330-360] & [-50,0] & 3511.29 \\

SDSS-NGC & [150,220] & [0,50] & 3072.38 \\

\hline

GAMA G09 & [129,141] & [-1,3] & 47.98 \\

GAMA G12 & [174,186] & [-2,2] & 47.99 \\

GAMA G15 & [211.5,223.5] & [-2,2] & 47.99 \\

\hline\hline
\end{tabular}
\caption{A summary of the main geometric properties of the Target fields used.}
\label{tb:fields}
 \end{center}
\end{table}

\section{Redshift Distributions}
\label{sec:nzdist}

\begin{figure}
\begin{center}
 \includegraphics[height=21.0cm]{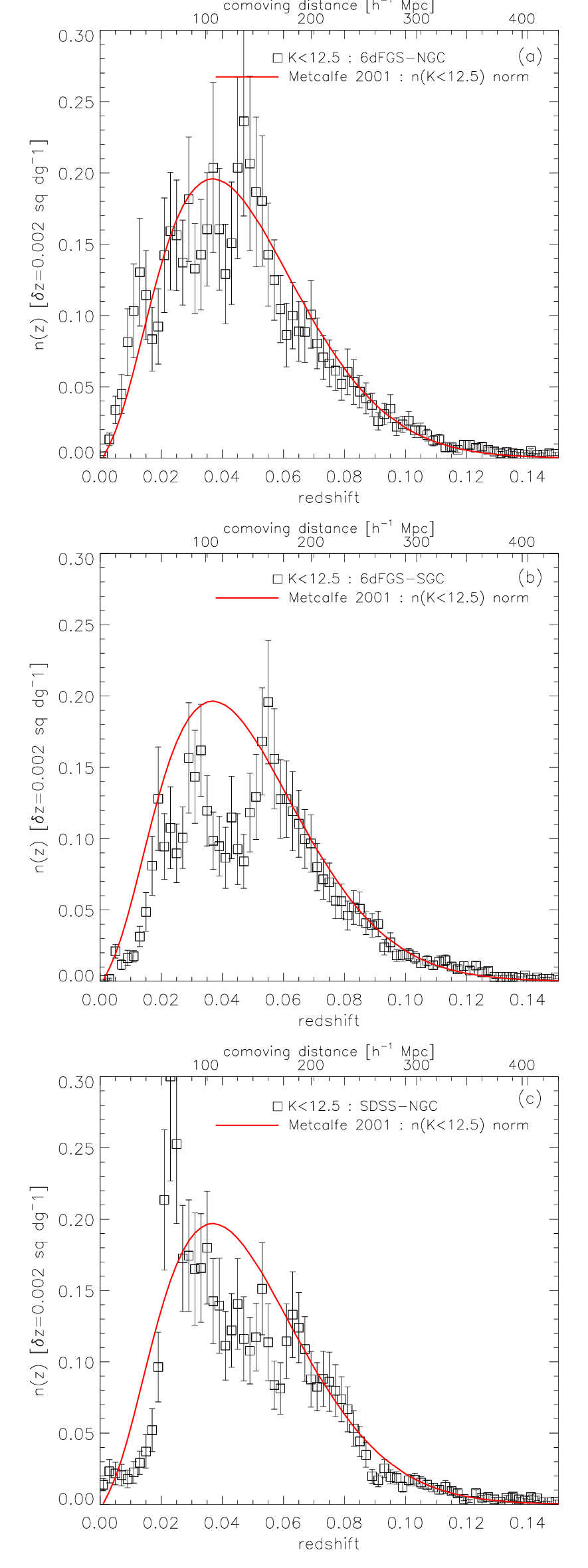}
 \caption{K band galaxy $n(z)$ with $K<12.5$ and $\delta z = 0.002$ normalised
using the $K < 12.5$ galaxy number counts. The red line represents the homogeneous
\citet{metcalfe_2001} LF prediction. The points (black, square) show data with
jack-knife derived errors.
\newline
 a) 6dFGS-NGC region (6dFGS, galactic north), \newline
 b) 6dFGS-SGC region (6dFGS, galactic south), \newline
 c) SDSS-NGC (SDSS$\otimes$2MASS, galactic north).}
 \label{fig:nz-ALL}
\end{center}
\end{figure}

\begin{figure}
\begin{center}
 \includegraphics[height=21.0cm]{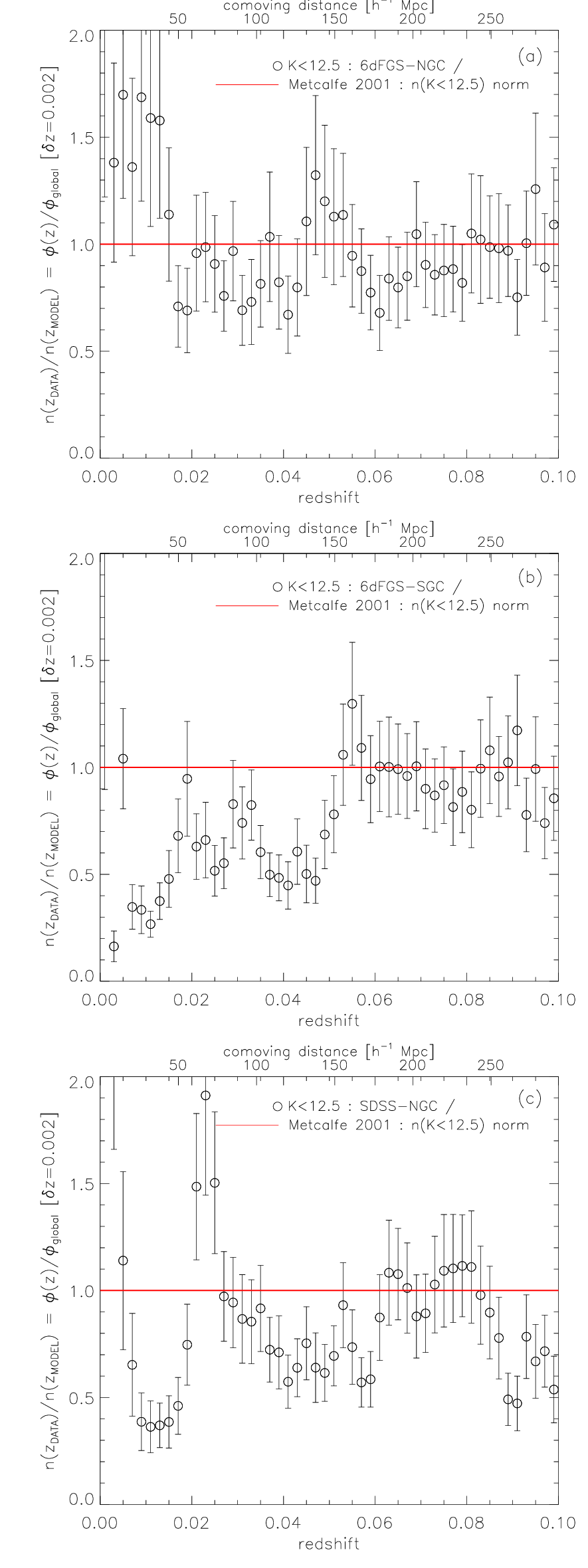}
 \caption{K band galaxy $\phi^*(z)/\phi_{\rm global}$ with $K < 12.5$ and
$\delta z = 0.002$ normalised using the $K < 12.5$ galaxy number counts. The red line
represents the homogeneous \citet{metcalfe_2001} LF prediction. The points
(black, circle) show data with jack-knife derived errors.
\newline
 a) 6dFGS-NGC region (6dFGS, galactic north), \newline
 b) 6dFGS-SGC region (6dFGS, galactic south), \newline
 c) SDSS-NGC (SDSS$\otimes$2MASS, galactic north).}
 \label{fig:nz-ALL-phiratio}
\end{center}
\end{figure}

We first probe the local galaxy clustering environment directly via
galaxy redshift distributions. Fig. \ref{fig:nz-ALL} shows the $n(z)$
distributions consistently limited at $K<12.5$ for our three target
regions. Here we are using 2MASS magnitudes matched to 6dFGS redshifts
in the case of 6dFGS-NGC and 6dFGS-SGC data and SDSS redshifts in the case
of SDSS-NGC. Errors have been estimated from jack-knife errors within
the 3 target regions. The red lines shows the homogeneous $n(z)$ model
estimated assuming the \citet{metcalfe_2001} LF and the K plus E corrections as
outlined in Section. \ref{sec:kecorrections}. These models
have been normalised so as to maintain the $K<12.5$ $n(m)$ underdensities stated
in Table \ref{tb:num_norm_table} and corrected for redshift incompleteness
(including any dependence of incompleteness on magnitude) using the method
described in Appendix \ref{append:magincompleteness}.

We then divided the observed $n(z)$ by this suitably normalised homogeneous
model to see over- and under-densities directly as a function of redshift. The
results are shown in Fig. \ref{fig:nz-ALL-phiratio} and the significant
non-uniformity we see reflects the presence of LSS in our local
universe. With this $K<12.5$ normalisation all three regions are typically
underdense for $z<0.05$ - see Table \ref{tb:num_norm_table}. The 6dFGS-SGC
region, which corresponds to the APM area \citep{maddox_1990}, is the most
underdense at $40\pm5$\%. The error here comes from jack-knife estimates. The
SDSS-NGC region is also significantly underdense at the $14\pm5$\% level. While the
6dFGS-NGC region still shows under-density, it is not significantly so
($4\pm10$\%). The error is bigger here because of the influence
of the Shapley-8 supercluster in this region. Therefore on scales out to
$\approx150$h$^{-1}$Mpc we conclude that the redshift distributions are
consistently underdense by $\approx4-40$\% with the South Galactic cap showing
the biggest under-density.

Clearly a lot depends on the accuracy of the $n(K)$ model normalisation.
\citet{frith_2006a} argued on the basis of a comparison of 2MASS $H<12.5$
magnitude counts to much fainter counts from Calar Alto OmegaCAM that
the \citet{metcalfe_2001} LF model normalisation  was supported by these
data. However, this count was only based on an area of 0.25deg$^2$. In
Section \ref{subsect:deepKcountsgama} we shall test if this normalisation is
consistent with the new $K$ band galaxy count data from the much bigger
150deg$^2$ area of the GAMA survey.

\begin{figure}
\begin{center}
 \includegraphics[height=21.0cm]{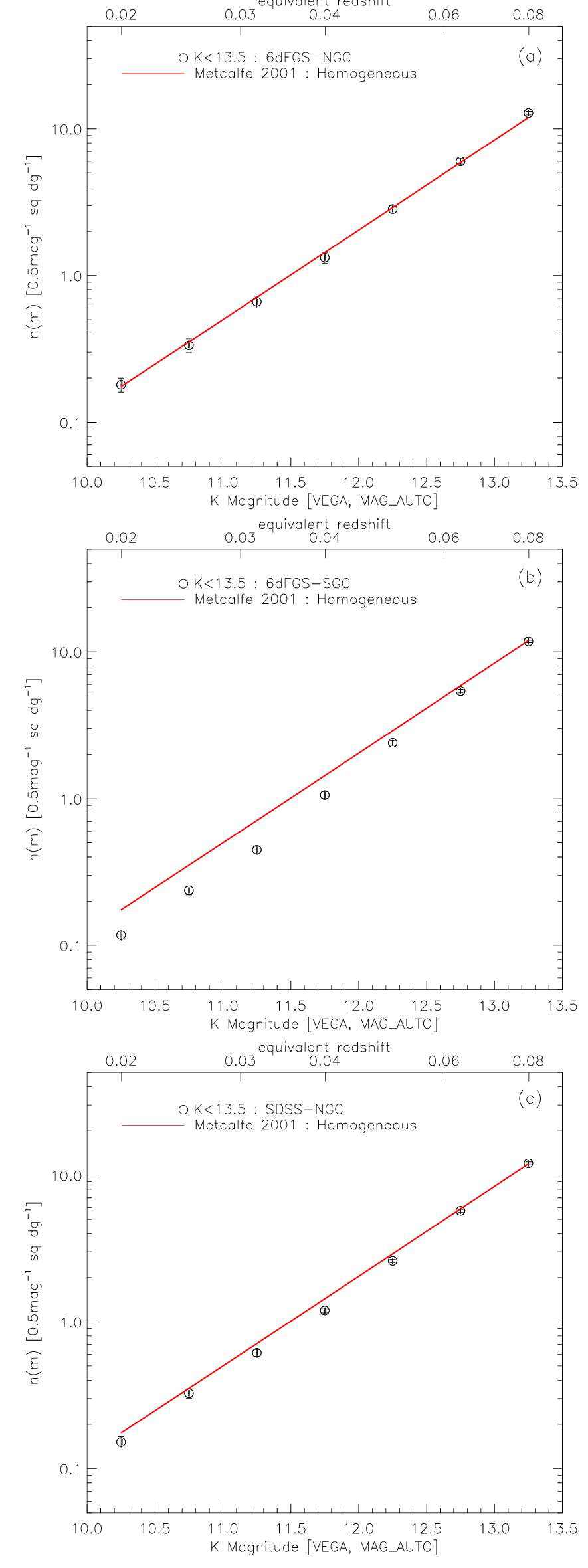}
 \caption{K band galaxy $n(m)$ from the 2MASS survey with $\delta m = 0.5$. The
red line represents the homogeneous \citet{metcalfe_2001} LF
prediction. The points (black, circle) show data with jack-knife derived errors.
\newline
 a) 6dFGS-NGC region (6dFGS, galactic north), \newline
 b) 6dFGS-SGC region (6dFGS, galactic south), \newline
 c) SDSS-NGC (SDSS$\otimes$2MASS, galactic north).}
 \label{fig:number-count-ALL}
\end{center}
\end{figure}

It is also still possible that a larger-scale under-density persists
beyond $z=0.05$ out to $z\approx0.1$. The underdensities then vary between
6-25\% as seen in Table \ref{tb:num_norm_table}. We find a weighted average
under-density of $15\pm3$\% for $K<12.5$ (with or without a $z<0.1$ cut).
Certainly a similar conclusion was reached by \citet{frith_2005a} who had the
advantage of the 2dFGRS $n(z)$ which reached fainter magnitudes and higher
redshifts but only covering a significantly smaller region of sky. Again, the
$n(z)$ model normalisation is even more crucial in measuring any under-density
at $0.05<z<0.1$ because a lot depends on the position of the homogeneous model
(red line) in Fig. \ref{fig:number-count-ALL}. This can be probed both by galaxy
counts to $K=15.8$ in the 150 deg$^2$ GAMA regions and $n(z)$ to $K=13.5$ by
virtue of the deeper redshift survey data in the  SDSS-NGC regions. But first we
return to check that our $n(z)$ results are consistent with the {\it form} of
the number counts to $K=13.5$.

\begin{table}
 \begin{center}
 \begin{tabular}{ccc}

\hline\hline
Field & Sample limit & Under-density \\
\hline\hline

6dFGS-NGC & $z<0.05$  & $0.96\pm0.10$ \\

6dFGS-SGC & $z<0.05$ & $0.60\pm0.05$ \\

SDSS-NGC & $z<0.05$ & $0.86\pm0.05$ \\

\hline

6dFGS-NGC & $z<0.1$  & $0.94\pm0.07$ \\

6dFGS-SGC & $z<0.1$ & $0.75\pm0.04$ \\

SDSS-NGC & $z<0.1$ & $0.86\pm0.04$ \\

\hline

6dFGS-NGC & $K<12.5$  & $0.96\pm0.07$ \\

6dFGS-SGC & $K<12.5$ & $0.76\pm0.03$ \\

SDSS-NGC & $K<12.5$ & $0.88\pm0.03$ \\

\hline

6dFGS-NGC & $K<13.5$  & $1.03\pm0.04$ \\

6dFGS-SGC & $K<13.5$ & $0.92\pm0.02$ \\

SDSS-NGC & $K<13.5$ & $0.97\pm0.02$ \\

\hline

SDSS-NGC & $r<17.2$ & $0.96\pm0.02$ \\

\hline\hline
\end{tabular}
\caption{A summary of the number count normalisations derived using the
homogeneous \citet{metcalfe_2001} LF prediction. These also correspond to 
under- and over-densities to the specified limits. The $z<0.05$ and $z<0.1$
entries assume $K<12.5$.
 }
\label{tb:num_norm_table}
 \end{center}
\end{table}

\begin{figure}
\begin{center}
 \includegraphics[height=21.0cm]{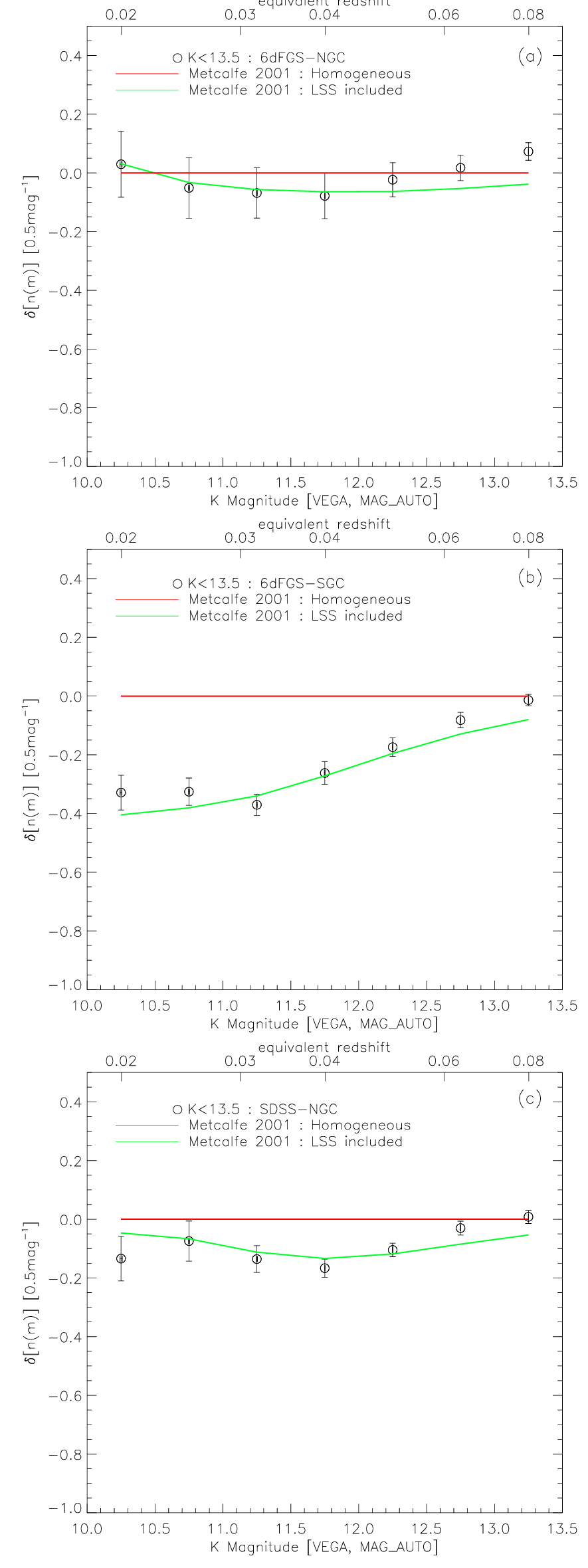}
 \caption{K band galaxy $n(m)$ density contrast from the 2MASS survey with
$\delta m = 0.5$. The red line represents the homogeneous \citet{metcalfe_2001}
LF prediction and the green line the LSS-corrected \citet{metcalfe_2001} LF
prediction. The points (black, circle) show data with jack-knife derived errors.
\newline
 a) 6dFGS-NGC region (6dFGS, galactic north), \newline
 b) 6dFGS-SGC region (6dFGS, galactic south), \newline
 c) SDSS-NGC (SDSS$\otimes$2MASS, galactic north).}
 \label{fig:number-fong-counts-ALL}
\end{center}
\end{figure}

\section{Number Counts}
\label{sec:nocounts}
\subsection{2MASS galaxy counts to \texorpdfstring{$K=13.5$}{K=13.5}}

Figs. \ref{fig:number-count-ALL} show the number counts to $K < 13.5$ for our 3
regions. In Appendix \ref{append:magaccuracy} we check for a scale-error in the
2MASS magnitudes and the statistics of star-galaxy separation as function of
magnitude. In fact, we do find a marginal scale error between $10<K<13.5$ and
all the magnitudes in Figs. \ref{fig:number-count-ALL} have been corrected for
this scale error. With or without this correction, all fields exhibit an
under-density relative to the homogeneous prediction (red line) until at least
$K\approx12.5$ and any convergence is only seen when the counts reach $K=13.5$. 

\noindent Using the $\phi^*(z)/\phi_{\rm global}$ correction for radial
inhomogeneity found earlier we show the LSS corrected model counts as 
the green line in Figs. \ref{fig:number-fong-counts-ALL} where observed counts
have been normalised by the homogeneous model. We see that accounting for the
inhomogeneities in the  $n(z)$ in Figs. \ref{fig:nz-ALL-phiratio} has improved
the model fit. This suggests a consistency between variations in the $n(z)$ and
$n(m)$ and a mutual agreement in the redshift under-density reported in Section
\ref{sec:nzdist}. 

These under-densities are either due to poor normalisation of the models at
fainter magnitudes, evolutionary brightening of galaxies at $z\approx0.1$ or
large-scale inhomogeneities. Note that the above scale error correction
tends to make the $K=13.5$ galaxy counts $\approx0.05$mag brighter,
slightly improving the fit to the homogeneous model. The generally improved
agreement between LSS corrected model and observed counts argues that the steep
number count slopes are not caused by systematics in the magnitudes or in
star-galaxy separation.

However, in all three regions the number counts are only becoming consistent
with homogeneity at  the $K = 13.5$ 2MASS survey limit, rather than the $K=12.5$
limit we used for the $n(z)$. This leaves the possibility open that the
under-density may extend beyond the scales we have used in our LSS corrections
and that the local volume remains underdense beyond $\approx150-300$h$^{-1}$Mpc.
We interpret the consistency between $n(m)$ and $n(z)$ as evidence for a local
hole-like under-density at least out to $z\approx0.08$.

\begin{figure*}
\begin{center}
 \includegraphics[width=10.cm]{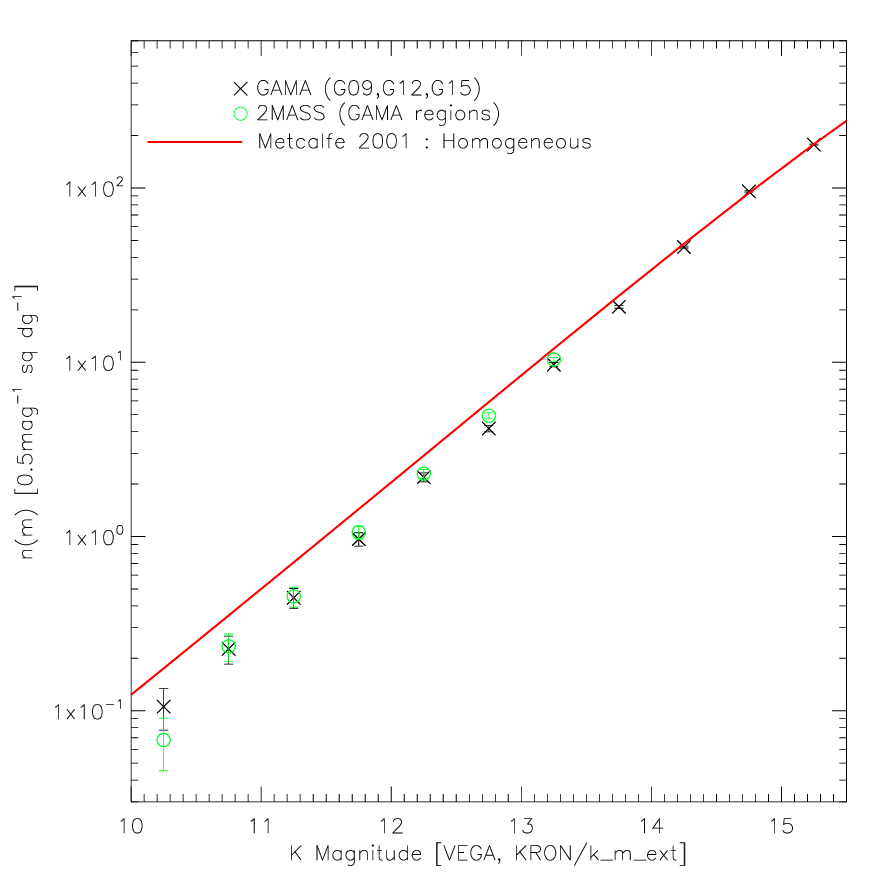}
 \caption{K band galaxy number counts comparing GAMA and 2MASS over
the GAMA regions. The red line represents the homogeneous \citet{metcalfe_2001}
LF prediction which at deep magnitudes is well normalised to the galaxy number
counts. The points show the 2MASS (green, circle) and GAMA (black, cross) data
with Poisson errors.
 }
 \label{fig:deepknumbercounts}
\end{center}
\end{figure*}

\subsection{Deeper \texorpdfstring{$K$}{K} counts from GAMA}
\label{subsect:deepKcountsgama}

We next use the GAMA survey over the full $3\times48$deg$^2$ regions
surveyed by the GAMA project to test the overall normalisation of the
homogeneous models for $n(z)$ and $n(m)$. We first calibrate the GAMA
$K$ Kron magnitudes to the 2MASS K $k\_m\_ext$ magnitude scale by
comparing the galaxy photometry. Using the `mpfitexy' routine we find that all
three GAMA regions are consistent with a one-to-one relation at
$\approx$1$\sigma$ as shown in Fig. \ref{fig:kk-gama-2mass}. However, again
using the `mpfitexy' routine we find and apply the $\approx -0.02$mag zeropoint
offsets detailed in Table \ref{tb:gamaukidss_2mass_koffsets}. We therefore
compare the GAMA K counts and the 3 GAMA fields of 2MASS galaxy counts to the
homogeneous models of \citet{metcalfe_2001} in Fig.
{\ref{fig:deepknumbercounts}. We see that the model fits the data well in the
range  $14<K<15.5$, supporting the normalisation we have used from Table
\ref{tb:shankslf}. We conclude that the normalisation we have used is reinforced
by the deeper K galaxy counts in the 150deg$^2$ of the GAMA region.

%
\section{\texorpdfstring{$N(\MakeLowercase{z})$}{N(z)} to
\texorpdfstring{$K=13.5$}{K=13.5} and
\texorpdfstring{$\MakeLowercase{r}=17.2$}{r=17.2} in the SDSS-NGC region}}

%
\begin{figure}
   \begin{minipage}{0.5\textwidth}
    \includegraphics[height=8.25cm]{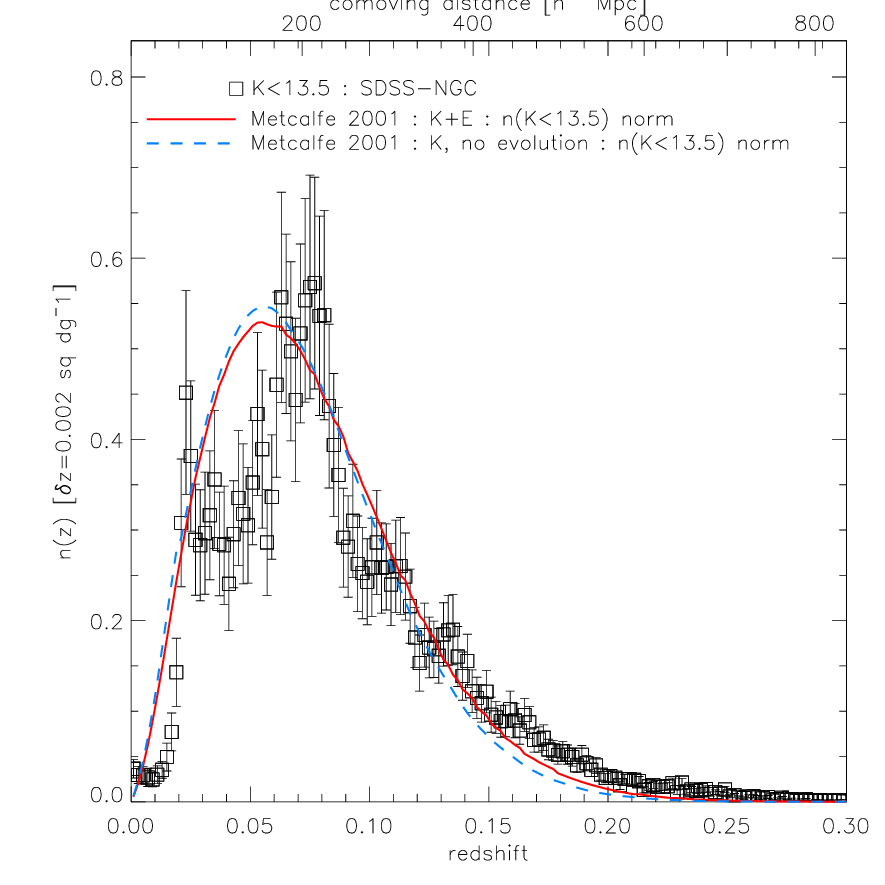}
    \caption{K band galaxy $n(z)$ with $K<13.5$ and $\delta z = 0.002$ normalised
using the $K < 13.5$ galaxy number counts. The red line represents the homogeneous
\citet{metcalfe_2001} LF prediction The points (black, square) show the SDSS-NGC data with
jack-knife derived errors.}
    \label{fig:numbercount-norm-nz}
    \end{minipage}
    \hspace{1pc}%
    \hfill
    \begin{minipage}{0.5\textwidth}
    \includegraphics[height=8.25cm]{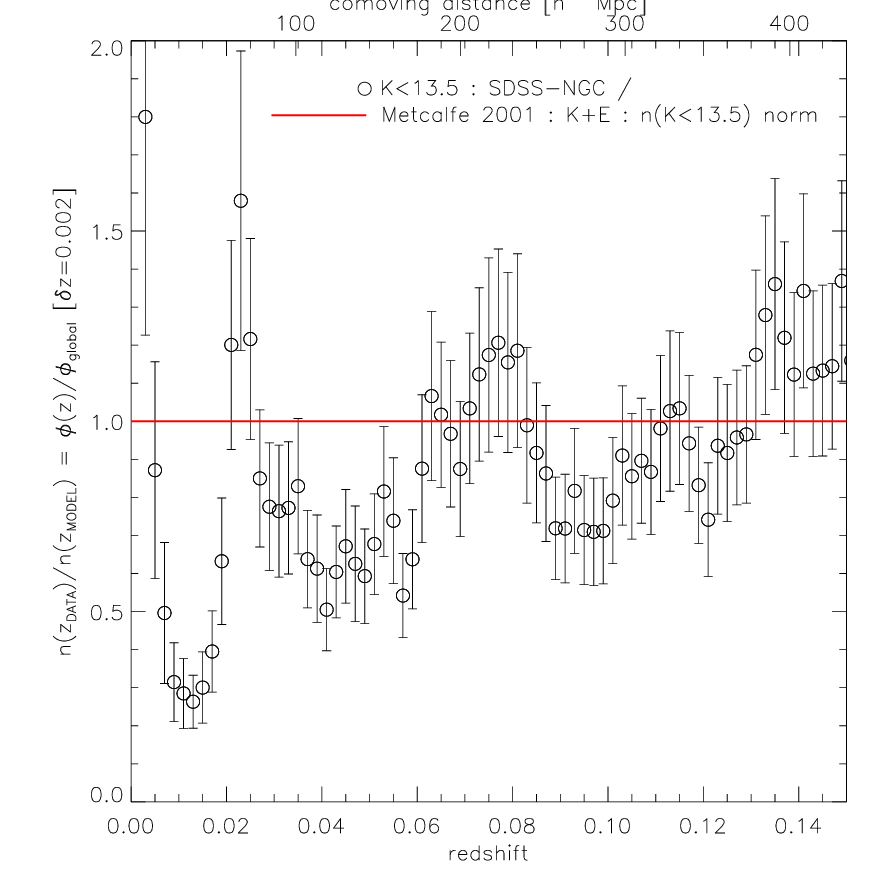}
    \caption{K band galaxy $\phi^*(z)/\phi_{\rm global}$ with $K < 13.5$ and
$\delta z = 0.002$ normalised using the $K < 13.5$ galaxy number counts. The red line
represents the homogeneous \citet{metcalfe_2001} LF prediction. The points
(black, circle) show the SDSS-NGC data with jack-knife derived errors.}
    \label{fig:numbercount-norm-phiz}
\end{minipage}
\end{figure}

\begin{figure}
   \begin{minipage}{0.5\textwidth}
    \includegraphics[height=8.25cm]{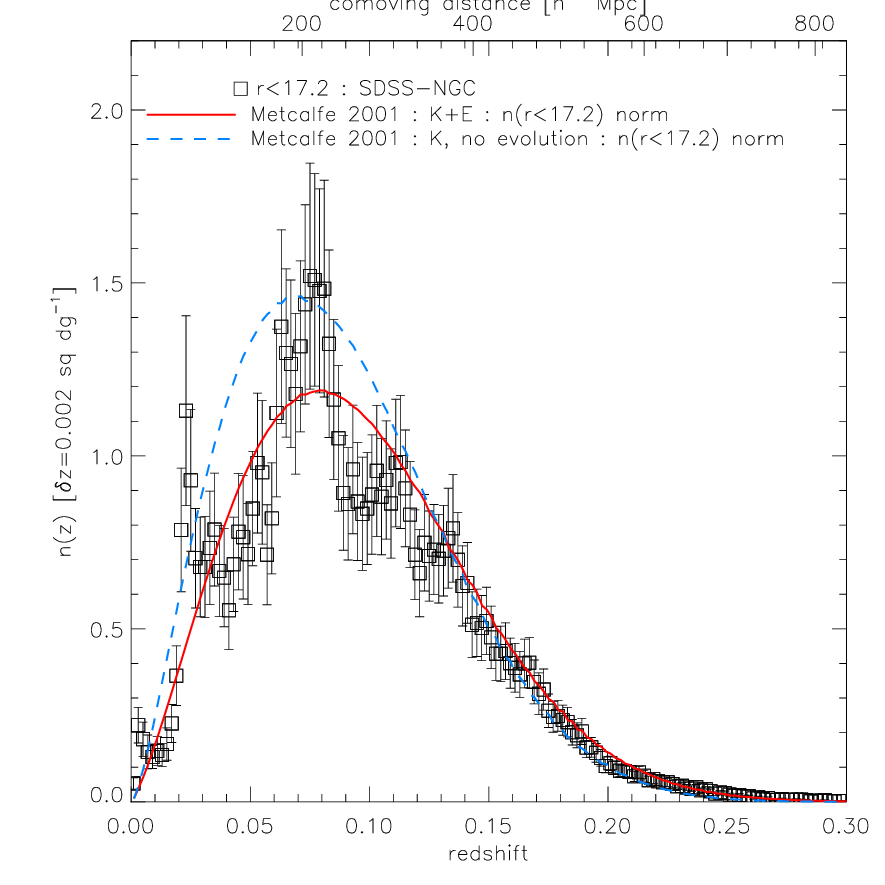}
 \caption{r band galaxy $n(z)$ with $r < 17.2$ and $\delta z = 0.002$ normalised
using the $r < 17.2$ galaxy number counts. The red line represents the
homogeneous \citet{metcalfe_2001} LF prediction. The points (black, square) show
the SDSS-NGC data with jack-knife derived errors.}
 \label{fig:optical-sdss-nz}
    \end{minipage}
    \hspace{5cm}
    \hfill%
    \begin{minipage}{0.5\textwidth}
    \includegraphics[height=8.25cm]{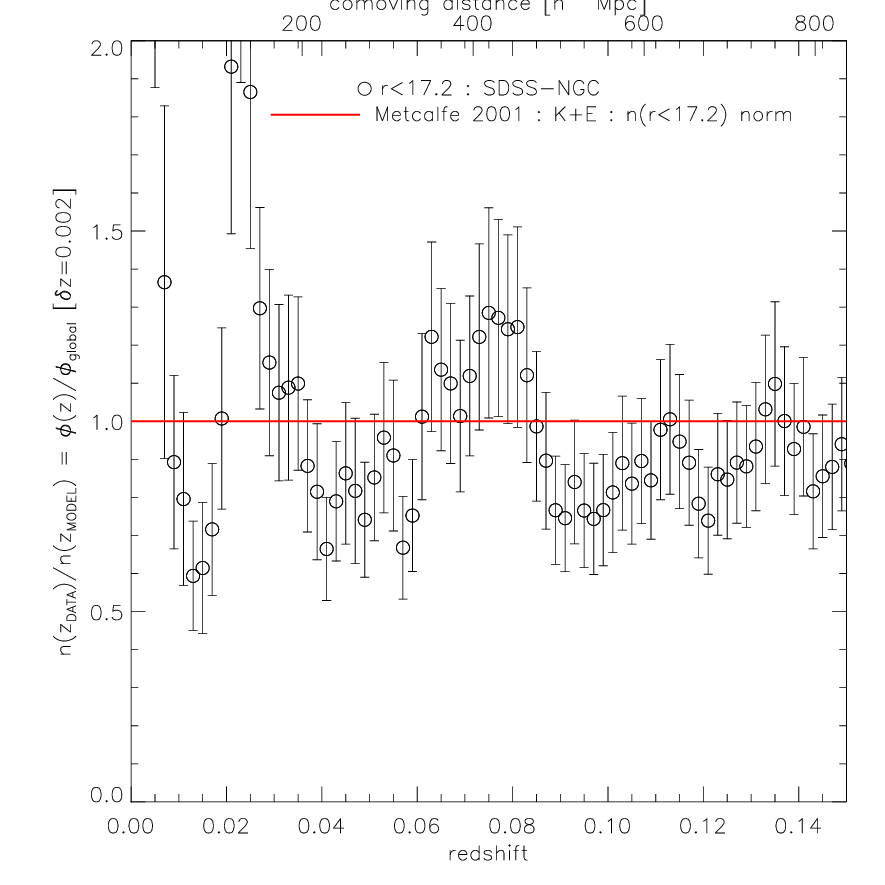}
 \caption{r band galaxy $\phi^*(z)/\phi_{\rm global}$ with $r < 17.2$ and
$\delta z = 0.002$ normalised using the $r < 17.2$ galaxy number counts. The red
line represents the homogeneous \citet{metcalfe_2001} LF prediction.  The points
(black, circle) show the SDSS-NGC data with jack-knife derived errors.}
 \label{fig:optical-sdss-phiz}
\end{minipage}
\end{figure}

It is possible to go to deeper z-survey limits in the SDSS-NGC region
because of the fainter magnitude limit in this region, compared to
6dFGS. Figs. \ref{fig:numbercount-norm-nz} and \ref{fig:numbercount-norm-phiz}
show the $n(z)$ and  $\phi^*(z)$ for this region to the $K=13.5$ limit of 2MASS.
We normalize the $n(z)$ by the 96\% ratio of data-model number magnitude counts
in this region to this limit - see Table \ref{tb:num_norm_table}. We note that
the same basic features in $n(z)$ are seen at low redshift but new over- and
under-densities appear at higher redshift. We note particularly the peak at
$z\approx0.08$. We see that it takes to $z\approx0.13$ before the model fits the
data.  Indeed, the K band counts in Fig. \ref{fig:number-count-ALL}(c) only
appear to converge at $K = 13.5$. We checked the difference that a no-evolution
model made to the $n(z)$ fit and it was small. The no-evolution $n(K)$ model is
also little different from the evolutionary model. The advantage of the $K$ band
is that it is less susceptible to evolutionary uncertainties.

Nevertheless, we also present the full $n(z)$ to $r=17.2$ in the
SDSS-NGC region. Here the $n(z)$ results are slightly  more ambiguous. The
$n(z)$ evolutionary model is compared to the data in Figs. \ref{fig:optical-sdss-nz} and
\ref{fig:optical-sdss-phiz}. The normalisation factor to $r<17.2$ from
the $n(r)$ is $0.96\pm0.02$. The $r<17.2$ $\phi(z)$  again shows evidence
for under-density but  here the observed $\phi^*(z)$  generally is
flatter, decreasing more slowly towards $z=0$ than in $K$. Also it shows
less indication of convergence at $z\approx0.1$.

\begin{figure*}
\begin{center}
 \includegraphics[width=10.cm]{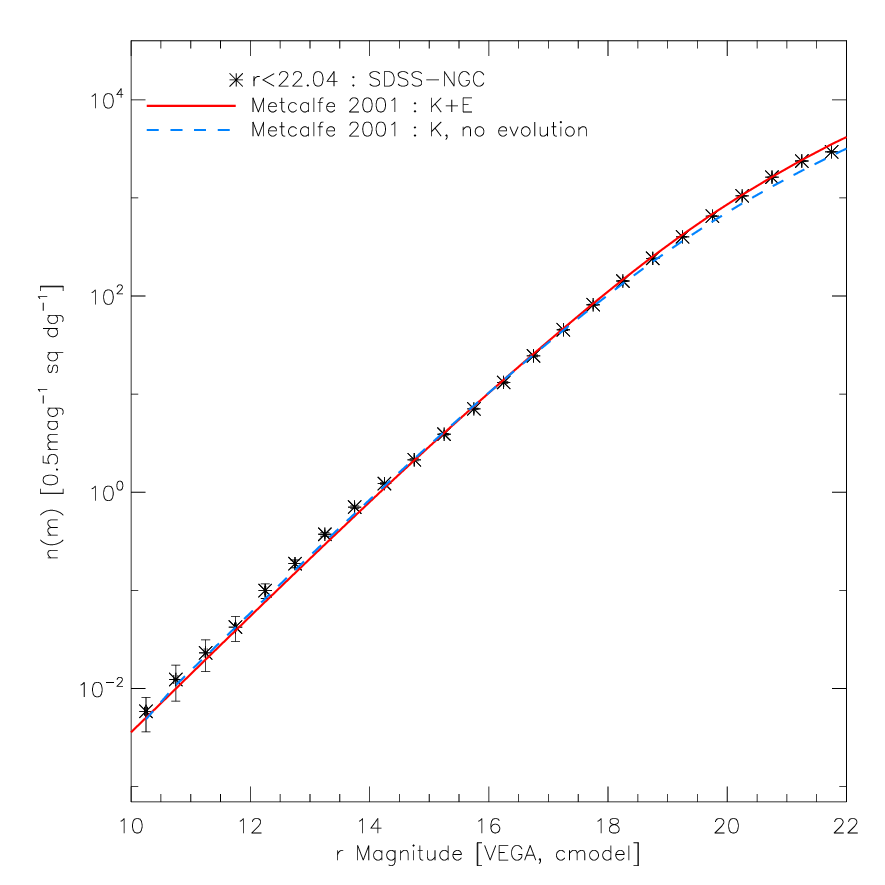}
 \caption{r band galaxy $n(m)$ with $\delta m = 0.5$. The red line represents
the homogeneous \citet{metcalfe_2001} LF prediction and the blue line the
no-evolution homogeneous \citet{metcalfe_2001} LF prediction. The points (black,
asterix) show the SDSS-NGC data with jack-knife derived errors.}
 \label{fig:optical-sdss-nm}
\end{center}
\end{figure*}

\begin{figure*}
\begin{center}
 \includegraphics[width=10.cm]{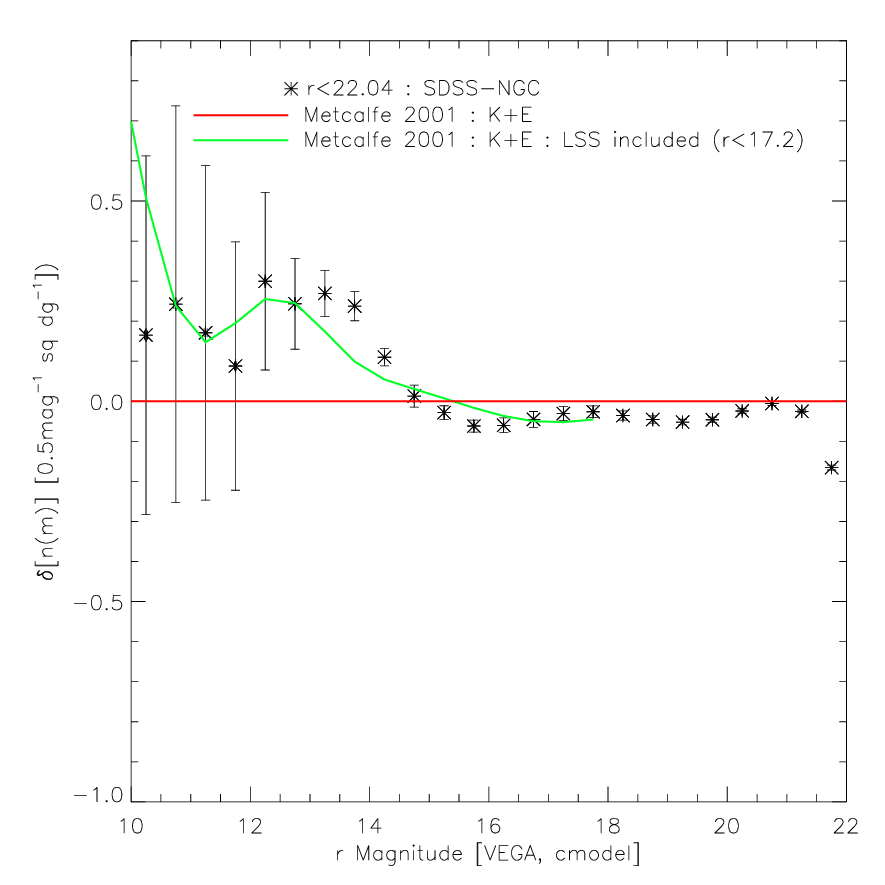}
 \caption{r band $n(m)$ based density contrast with $\delta m = 0.5$. The red line
represents the homogeneous \citet{metcalfe_2001} LF prediction and the green
line the LSS-corrected \citet{metcalfe_2001} LF prediction. The points (black,
asterix) show the SDSS-NGC data with jack-knife derived errors.}
 \label{fig:optical-sdss-nm-euclidean}
\end{center}
\end{figure*}

Clearly the normalising factor inferred from the r-band count is crucial here
and we show $n(r)$ to $r<22$ for the SDSS-NGC region in
Figs. \ref{fig:optical-sdss-nm} and \ref{fig:optical-sdss-nm-euclidean}. These
counts are consistent with the \citet{yasuda_2001} analysis of the
SDSS commissioning data for the magnitude range $15<r<20$. A
similar behaviour is seen in Fig. \ref{fig:optical-sdss-nm-euclidean} as in
Fig. \ref{fig:optical-sdss-phiz} in that the observed $n(r)$ takes till
$r\approx20$ to reach the homogeneous model. This is reinforced by the
approximate agreement of the counts with LSS corrected model based on the
$r<17.2$ $n(z)$. Thus there is at least consistency  between the suggestions
from $n(z)$ and $n(m)$ for the under-density extending beyond $z=0.1$.

Furthermore, there is uncertainty caused by the increased possibility of
evolution in the $r$ band. A no-evolution model for $n(m)$ is therefore
also shown in Fig. \ref{fig:optical-sdss-nm}. This model has a flatter slope and
therefore reaches agreement with $n(r)$ at a brighter $r=19$
magnitude. Thus here there would both be stronger evidence for a void
within say 150h$^{-1}$Mpc but the evidence for a more extended
under-density would be less than with the evolutionary model. It should
also be noted that within the classes of models considered here, an
evolutionary model gives a better fit to $n(r)$ at $r>20$. 

Uncertainties in the count normalisation and the evolutionary model thus
appear to be  more significant in the $r$ band  and this reinforces the
advantage of working in $K$. The $K$ band counts may also be more
sensitive to over- and under-densities, being more dominated by strongly
clustered early-type galaxies. We conclude that the evidence in the K
band for a local hole out to 300h$^{-1}$ Mpc can be regarded as more
reliable than the more ambiguous evidence for a flatter under-density to
greater distances from the $r<17.2$ $n(z)$.

%

\clearpage

\section{The Hubble Diagram}
\label{sec:hubblediaresults}

\begin{figure}
\begin{center}
 \includegraphics[height=21.0cm]{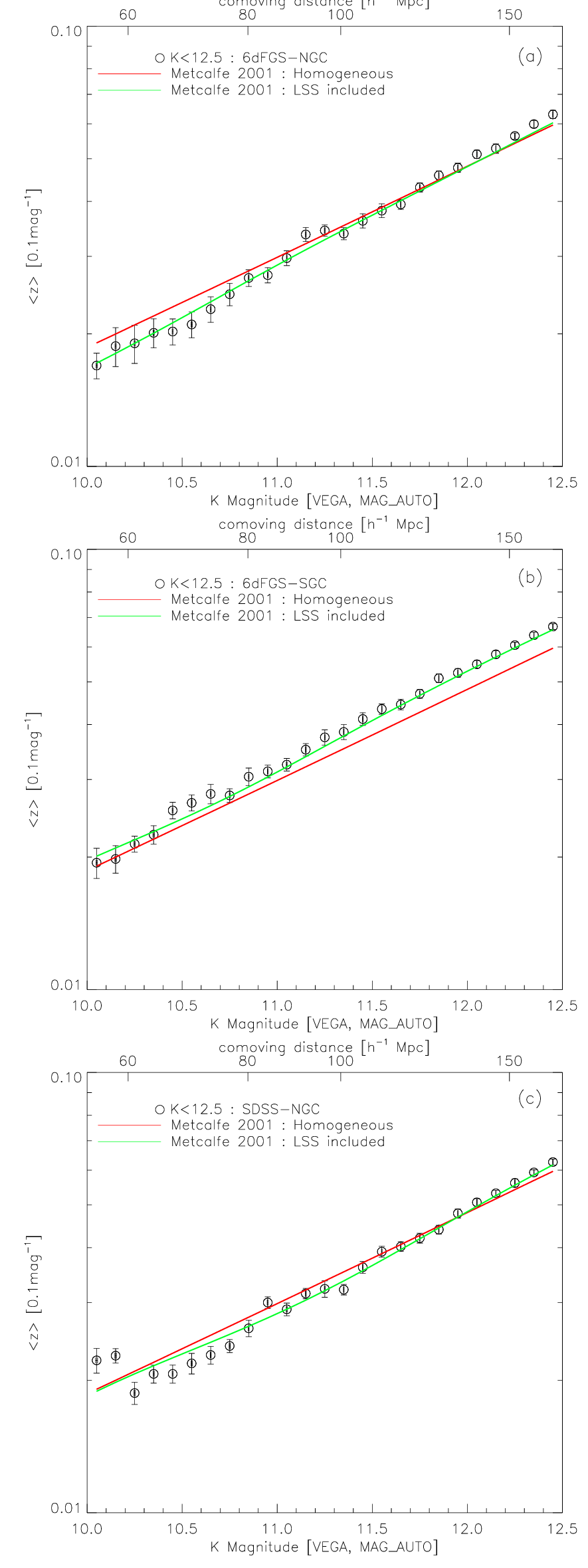}
 \caption{K band $\overline{z}(m)$ with $\delta m = 0.1$. The red line
represents the homogeneous \citet{metcalfe_2001} LF prediction and the green
line the LSS-corrected \citet{metcalfe_2001} LF prediction. The points (black,
circle) show data with jack-knife derived errors. \newline
 a) 6dFGS-NGC region (6dFGS, galactic north), \newline
 b) 6dFGS-SGC region (6dFGS, galactic south), \newline
 c) SDSS-NGC (SDSS$\otimes$2MASS, galactic north).
 }
 \label{fig:zbar-ALL}
\end{center}
\end{figure}

\begin{figure}
\begin{center}
 \includegraphics[height=21.0cm]{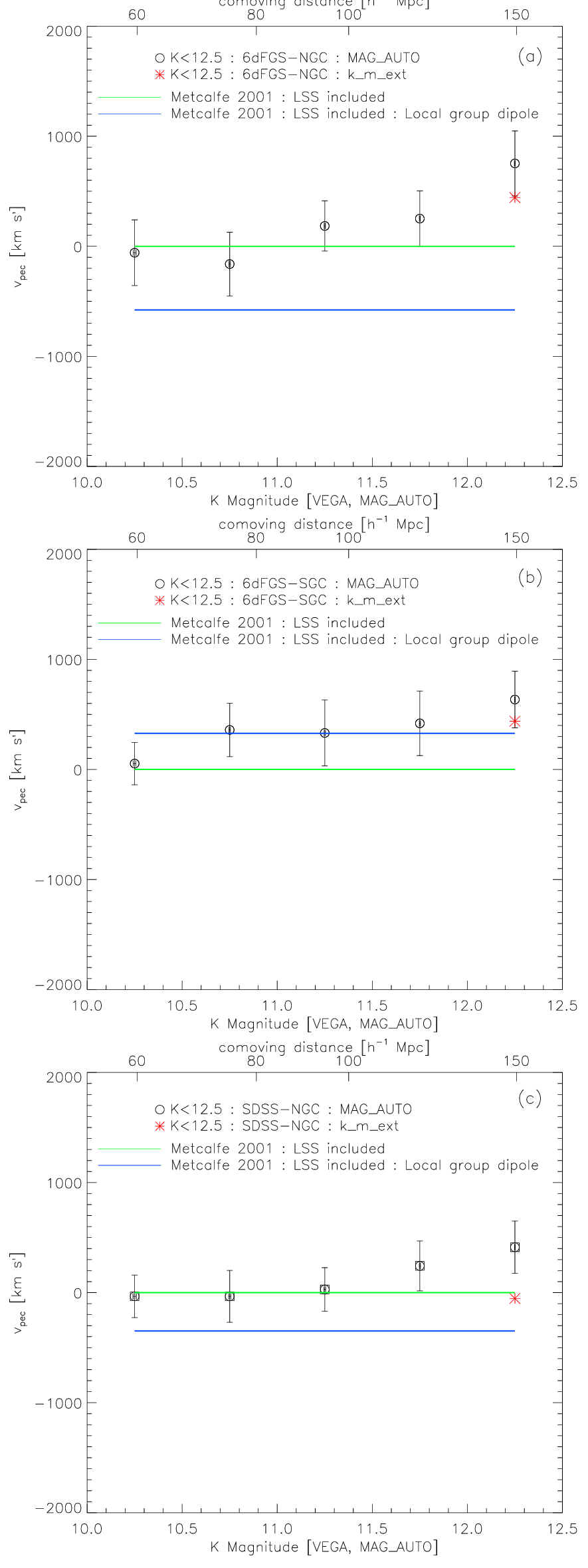}
 \caption{K band $\overline{z}(m)$ with $\delta m = 0.5$. The green
line represents the LSS-corrected \citet{metcalfe_2001} LF prediction and the
blue line the CMB dipole flow LSS-corrected \citet{metcalfe_2001} LF prediction.
The points (black, circle) show data with jack-knife derived errors. The red asterisk
shows the final bin without corrections.\newline
 a) 6dFGS-NGC region (6dFGS, galactic north), \newline
 b) 6dFGS-SGC region (6dFGS, galactic south), \newline
 c) SDSS-NGC (SDSS$\otimes$2MASS, galactic north).
 }
 \label{fig:zbar-tomresidual-ALL-halfbin}
\end{center}
\end{figure}

Fig. {\ref{fig:zbar-ALL}} shows $\overline{z}(m)$ for our three fields.
The homogeneous prediction for each region is shown as the red
line and the LSS corrected model, based on  the $\phi^*(z)$ found earlier,
is shown as the green line. In all three cases we see that the
green line gives an improved, although not perfect, fit to the data. But
the importance of the LSS correction is clear since the under-prediction of the
observed $\overline{z}(m)$ particularly in the 6dFGS-SGC region might otherwise
be interpreted as immediately implying peculiar motion which is clearly not the
case. As expected, the 6dFGS-SGC region shows the biggest LSS (red-green)
correction between the 3 regions since it showed the biggest low-redshift
under-density in Fig. {\ref{fig:nz-ALL-phiratio}} but the other two regions also
tend to behave similarly. We also note the tentative `spike' in $\overline{z}(m)$
in the 6dFGS-NGC region at $K\approx11.5$, which is the approximate
location of the Shapley-8 supercluster. But even with $\delta m=0.1$mag
bins, this technique does not have the resolution to detect backside
infall etc.

To examine these $\overline{z}(m)$ relations in more detail, we next
subtract the LSS corrected `Hubble law' prediction from the data in
Fig. \ref{fig:zbar-ALL}. This means we are in effect plotting a sky-averaged
$v_{pec}$. The results are shown in Fig.
\ref{fig:zbar-tomresidual-ALL-halfbin} for a magnitude bin of $\delta m = 0.5$.
For comparison purposes we also show the $\overline{z}(m)$ for the final
$K=12.25$ bin when using the original 2MASS `k\_m\_ext' magnitude and without
the spectroscopic completeness corrections described in Appendix
\ref{append:magincompleteness}. The difference between these results means we
cannot place too much weight on this final bin when interpreting these data.
However, we note that over the rest of the magnitude range this difference
is small, particular so for the 6dFGS-NGC/SGC fields. We therefore conclude
that completeness corrections are only important for the final magnitude bin
and the SDSS-NGC data.

Since the models indicated by the green lines assume galaxies are at
rest in the Local Group frame then this is tantamount to assuming that
all galaxies and the Local Group are moving with a coherent bulk notion. We now
investigate an alternative hypothesis that the Local Group is moving with
633kms$^{-1}$ relative to more distant galaxies i.e: the CMB dipole motion in
the Local Group frame. The relative average recession velocity of these distant
galaxies should then be correspondingly reduced in the direction of our motion
relative to the CMB and increased in the opposite direction. This `dipole'
non-bulk motion model is then represented by the blue lines in Figs.
\ref{fig:zbar-tomresidual-ALL-halfbin}.

We immediately see that in two out of three regions the bulk motion
prediction agrees with the data much better than the non-bulk motion model where
only the Local Group is moving with 633kms$^{-1}$ with respect to the CMB. Even
in the third region in the 6dFGS SGC direction, although the data agrees
better with the non-bulk motion model, it is also still in reasonable
agreement with the bulk motion model. The significance of the rejection of the
non-bulk motion model has been estimated using the $K=11.75$ bin. This is
necessary as the smoothing by the galaxy luminosity function causes different
magnitude bins to be highly covariant and also the final bin may be less
reliable as discussed above. The level of rejection of the non-bulk motion model
in the 6dFGS-NGC and SDSS-NGC regions is at the $3.1\sigma$ and $2.3\sigma$
levels respectively. This suggests that at least in the 6dFGS-NGC and SDSS-NGC
directions we may be seeing a bulk motion with convergence to the CMB dipole not
yet reached at our $K<12.5$mag survey limits. Combining the measurements across
all three regions we find a overall rejection of the non-bulk motion model at
the $3.9\sigma$ level. In contrast, the bulk motion model is consistent with the
data overall at the $1.5\sigma$ level. The fit of the bulk motion model
indicates that the scale of convergence is larger than the
$\approx150$h$^{-1}$Mpc scale probed at $K<12.5$. However, it should be noted
that the residual dipole effect is small relative to the LSS correction.

It is somewhat counter-intuitive that the regions which are less underdense on
average (6dFGS-NGC, SDSS-NGC) agree with the bulk motion model whilst the most
underdense region (6dFGS-SGC) agrees with the dipole based non-bulk motion
model. However, this might be consistent with a faster local expansion in the
most underdense area. In this view the agreement of 6dFGS-SGC $\overline{z}(m)$
with the non-bulk motion model (blue line) would be accidental with the real
interpretation being a bulk motion (green line) combined with a faster local
expansion resulting in an excess $v_{pec}$ as is observed. We note that in the
other two regions there is at least no inconsistency with a faster local
expansion rate relative to the bulk motion model. But it should still be noted
that our simple models do not include peculiar velocities generated by
structures like Shapley-8 in 6dFGS-NGC which would produce apparently higher
expansion rates even beyond their nominal redshift, due to the smoothing of
$\overline{z}(m)$ by the galaxy luminosity function. Similarly, these models
may be somewhat affected by inhomogenous Malmquist bias from LSS at
deeper redshifts that is not fully accounted for with our $K<12.5$ derived
$\phi(z)$ density profiles. 

We conclude that the successful fit of a bulk motion model fit to
$\overline{z}(m)$ may be consistent with the $\approx$150h$^{-1}$Mpc scale
coherent under-density found in $n(z)$ and $n(m)$ across our three regions. The
question of whether the 300h$^{-1}$Mpc void is visible dynamically in
$\overline{z}(m)$ is less clear because that statistic does not reach
$z\approx0.1$. Clearly the SNIa Hubble diagram probes out to larger redshifts
where it is a more probable standard candle than our galaxy samples. The
question then of whether there is dynamical evidence of a Local Hole is of
course intertwined with the cosmological model that is assumed.

\section {Conclusions}

We have used $n(m)$ from 2MASS and $n(z)$ from 6dFGS and SDSS limited at
$K<12.5$ over much of the  sky at high galactic latitudes to probe the
local large-scale  structure, extending the work of \citet{frith_2005b}. We
looked at three volumes and found that that in the 6dFGS-SGC region, which
broadly corresponds to the area previously covered by the APM survey,
there is a clear $\approx40$\% under-density out to 150h$^{-1}$Mpc. In the
SDSS-NGC volume an $\approx15$\% under-density is seen
again out to 150h$^{-1}$Mpc although this is broken by the Coma cluster
producing a strong over-density at $\approx75$h$^{-1}$Mpc  in front of large
under-densities behind it. A $\approx5$\% under-density is seen in the 6dFGS-NGC
area out to about 150h$^{-1}$Mpc.  The implied local under-density in $n(z)$ and
$n(m)$ averaged over the 3 fields out to $K<12.5$ is $\approx15\pm3$\%.
Modelling the $K$ number counts using the ratio of a homogeneous model
normalised to these over- and under-densities to define
$\phi^*(z)$, produced good agreement with the under-densities seen in the number
counts to $K=12.5$, particularly in the 6dFGS-SGC area. This agreement between
$n(m)$ and $n(z)$  supports the reality of these local inhomogeneities
out to $\approx150$h$^{-1}$Mpc depth.

While $\Lambda$CDM may allow structures on 200-300h$^{-1}$Mpc scales
\citep{yadav_2010,park_2012,watson_2013},
\citet{frith_2005b} calculated that a 24\% under-density to $K<12.5$
over the 4000deg$^2$ APM (6dFGS-SGC) area would be inconsistent with the
$\Lambda$CDM model at the 4-4.5$\sigma$ level, depending on whether the
calculation was based on a theoretical $\Lambda$CDM or  observed 2MASS
$w(\theta)$ - see their Table 1. However, when these authors take into account
the previous uncertainties in the K band count normalisation, this significance then
reduced to 2-3$\sigma$. Here, we have confirmed the 6dFGS-SGC
under-density to be $24\pm3$\% at $K<12.5$ in only a slightly smaller area
(3511deg$^2$) and further confirmed that our number count normalisation
is accurate from the deeper GAMA data, in an area $\approx600\times$ larger than
that available to \citet{frith_2005b}. So the existence of such a
coherent under-density in the South Galactic cap appears to imply an
$\approx4\sigma$ discrepancy with the $\Lambda$CDM model, in terms of the
large-scale power that it predicts. 

The use of the luminosity function of \citet{metcalfe_2001}  is a
potential area of weakness in these studies. However,
\citet{shanks_2013} use  maximum likelihood techniques to estimate the
luminosity function and $\phi^*(z)$ simultaneously for the $r$ and $K$ limited
samples. They find that our assumed luminosity function is either
in good agreement with the self-consistently estimated luminosity
function ($r$-band) or where it differs slightly ($K$-band) the
$\phi^*(z)$ results prove robust and unaffected.

We then made a Hubble diagram using the $\overline{z}(m)$ technique of
\citet{soneira_1979}. Before we could detect peculiar velocities  we had to
make LSS corrections to make the model for $\overline{z}(m)$ take
account of the inhomogeneities already found. In the 6dFGS-SGC region we
found that the LSS-corrected $\overline{z}(m)$ prefers a solution that
includes a 633kms$^{-1}$ CMB velocity component for the Local Group
relative to galaxies in this direction. In the 6dFGS-NGC and SDSS-NGC regions
the more distant galaxies still preferred the solution without the CMB velocity
added to the Local Group and so can be said to prefer a bulk motion
solution where the local motion towards the CMB dipole direction has
not converged.

The local under-densities we have found will imply faster local
expansions. Indeed, we noted that such a scenario is not inconsistent
with the results we found with $\overline{z}(m)$. Such a faster local
expansion could help alleviate the tension at the $\approx5$\% level
between recent local and CMB measures of $H_{0}$
\citep{planckpara_2013}. The naive expectation for the effect on $H_{0}$
can be derived by assuming linear theory, $\delta H_{0}/H_{0} =
- \tfrac{1}{3} \Omega_m^{0.6}/b \times \delta \rho_{g}/\rho_{g}$ where the
bias,  $b\approx1$, for the standard model. Then the $19\pm3$\%, $z<0.05$,
$K<12.5$, under-density we report suggests an $\approx2-3$\% increase in
$H_{0}$. Indeed, this level of variation is not inconsistent with estimates of
the cosmic variance of $H_{0}$ in $\Lambda CDM$ \citep{kalus_2013,marra_2013}.
However, for the Southern Galactic cap region where we found a deeper
underdensity of $\approx40$\%, a larger $H_{0}$ correction of $6-7$\% would be
implied.

Finally, we investigated the evidence for an even larger local under-density
out to $\approx300$h$^{-1}$Mpc. We first determined the $n(m)$
normalisation at fainter $K\approx16$mag and $r\approx20.5$mag from GAMA
and SDSS. We found excellent agreement with the K model counts at
$K\approx15$. This normalisation implies that the under-density in the
SDSS-NGC volume may extend to $\approx300$h$^{-1}$Mpc and even deeper if
the SDSS-NGC $r<17.2$ $n(z)$ is to be believed. However, there is 
increased uncertainty in $r$ due to the likelihood of increased
evolutionary effects as well as the count model normalisation
uncertainty. Although $\overline{z}(m)$ at these limits cannot  test further this
300h$^{-1}$Mpc under-density dynamically, we have noted that any
cosmology that fits the SNIa Hubble diagram before accounting for the
Local Hole must fail at some level afterwards.

\section*{Acknowledgments}

We thank Richard Fong, Michael Hill, John Lucey, Utane Sawangwit and Maciej
Bilicki for useful comments. JRW acknowledges financial support from STFC. We
also thank the anonymous referee for their useful comments.

Funding for SDSS-III has been provided by the Alfred P. Sloan Foundation, the
Participating Institutions, the National Science Foundation, and the U.S.
Department of Energy Office of Science. The SDSS-III web site is
http://www.sdss3.org/.

SDSS-III is managed by the Astrophysical Research Consortium for the
Participating Institutions of the SDSS-III Collaboration including the
University of Arizona, the Brazilian Participation Group, Brookhaven National
Laboratory, University of Cambridge, Carnegie Mellon University, University of
Florida, the French Participation Group, the German Participation Group, Harvard
University, the Instituto de Astrofisica de Canarias, the Michigan State/Notre
Dame/JINA Participation Group, Johns Hopkins University, Lawrence Berkeley
National Laboratory, Max Planck Institute for Astrophysics, Max Planck Institute
for Extraterrestrial Physics, New Mexico State University, New York University,
Ohio State University, Pennsylvania State University, University of Portsmouth,
Princeton University, the Spanish Participation Group, University of Tokyo,
University of Utah, Vanderbilt University, University of Virginia, University of
Washington, and Yale University.

This publication makes use of data products from the Two Micron All Sky Survey,
which is a joint project of the University of Massachusetts and the Infrared
Processing and Analysis Center/California Institute of Technology, funded by the
National Aeronautics and Space Administration and the National Science
Foundation

GAMA is a joint European-Australasian project based around a spectroscopic
campaign using the Anglo-Australian Telescope. The GAMA input catalogue is based
on data taken from the Sloan Digital Sky Survey and the UKIRT Infrared Deep Sky
Survey. Complementary imaging of the GAMA regions is being obtained by a number
of independent survey programs including GALEX MIS, VST KiDS, VISTA VIKING,
WISE, Herschel-ATLAS, GMRT and ASKAP providing UV to radio coverage. GAMA is
funded by the STFC (UK), the ARC (Australia), the AAO, and the participating
institutions. The GAMA website is http://www.gama-survey.org/ 

This research has made use of the NASA/IPAC Extragalactic Database (NED) which
is operated by the Jet Propulsion Laboratory, California Institute of
Technology, under contract with the National Aeronautics and Space
Administration. This research has made use of the VizieR catalogue access tool,
CDS, Strasbourg, France''. The original description of the VizieR service was
published in A\&AS 143, 23 (2000). We have made use of the MPFITEXY routine
\citep{williams_2010}. The MPFITEXY routine depends on the MPFIT
package \citep{markwardt_2009}. We would also like to acknowledge the use of the
TOPCAT utility.


\setlength{\bibhang}{2.0em}
\setlength\labelwidth{0.0em}
%

\newpage
\appendix

\section{Magnitude Accuracy}
\label{append:magaccuracy}
\subsection{2MASS k\_m\_ext}

\begin{figure}
\begin{center}
 \includegraphics[height=7.25cm]{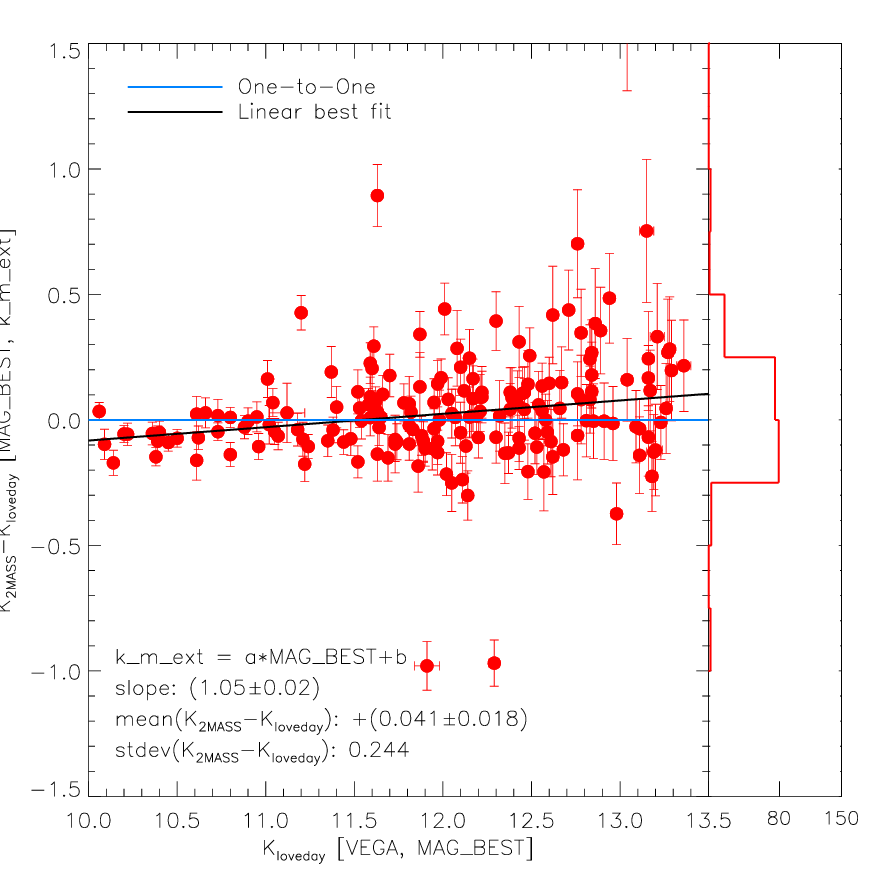}
 \caption{K band magnitude comparison for 181 common galaxies of the deep K data
of \citet{loveday_2000} who have provided the $MAG\_BEST$ magnitude from
SeXtractor to the corresponding 2MASS k\_m\_ext magnitude. The derived slope
using the `mpfitexy' routine and both the mean and standard deviation of the
residuals are stated.
}
 \label{fig:kcalloveday}
\end{center}
\end{figure}

Here we aim to test for scale and zeropoint errors in our 2MASS k\_m\_ext
magnitudes. We therefore compare to the previous galaxy photometry of
\citet{loveday_2000} where pseudo-total MAG\_BEST magnitudes were measured using
SeXtractor. In Fig. \ref{fig:kcalloveday} we show the resulting comparison after
matching the \citet{loveday_2000} galaxies to 2MASS with a $3''$ matching
radius.

First, assuming no scale error we find a marginally significant zeropoint
offset of k\_m\_ext-MAG\_BEST=$0.04\pm0.02$mag. Then we test for non-linearity by
fitting for a scale-error using the `mpfitexy' routine considering errors in
both magnitudes. We find a slope of k\_m\_ext = ($1.05\pm0.02$)MAG\_BEST.
Whilst this is only significant at the $\approx 2 \sigma$ level, we
nevertheless applied this correction factor to the k\_m\_ext magnitude thereby
placing the 2MASS data on the \citet{loveday_2000} system. Although this has the
effect of slightly steepening the 2MASS counts in Fig.
\ref{fig:number-count-ALL} the effect on the overall conclusions is
negligble.

\begin{figure}
\begin{center}
 \includegraphics[height=7.cm]{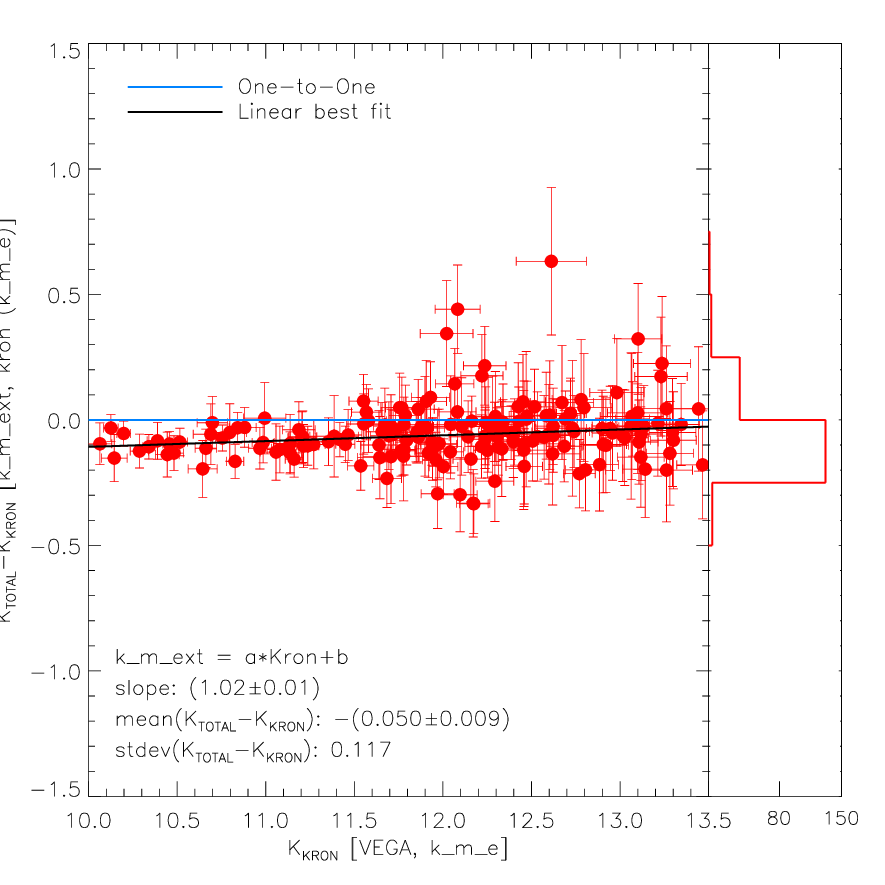}
 \caption{An internal K band magnitude comparison of the 2MASS k\_m\_ext and the
elliptical Kron (k\_m\_e) magnitudes for 181 common galaxies of the deep K data
of \citet{loveday_2000}. The derived slope using the `mpfitexy' routine and both
the mean and standard deviation of the residuals are stated.
}
 \label{fig:kcal2mass}
\end{center}
\end{figure}

We further check the 2MASS k\_m\_ext magnitude by comparing to the 2MASS Kron
magnitude (k\_m\_e) for the \citet{loveday_2000} galaxies in Fig. \ref{fig:kcal2mass}. Both these 2MASS
magnitudes are pseudo-total and so a one-to-one relationship might be expected.
First we find a simple offset of k\_m\_ext-Kron=$-0.05\pm0.01$mag. Although this is
significant, for this work offsets are less important than scale errors. We test
for such a scale error as above and find a slope of k\_m\_ext =
($1.02\pm0.01$)Kron thus the k\_m\_ext and Kron magnitudes seem reasonably
consistent with a one-to-one relation.

We note that in Fig. \ref{fig:deepknumbercounts} we have not corrected the
2MASS+GAMA magnitudes onto the \citet{loveday_2000} system. This is conservative
since the effect would be to imply a slightly higher ($\approx3$\%)
normalisation for our \citet{metcalfe_2001} LF and homogeneous counts model.

\begin{figure}
\begin{center}
 \includegraphics[height=7.25cm]{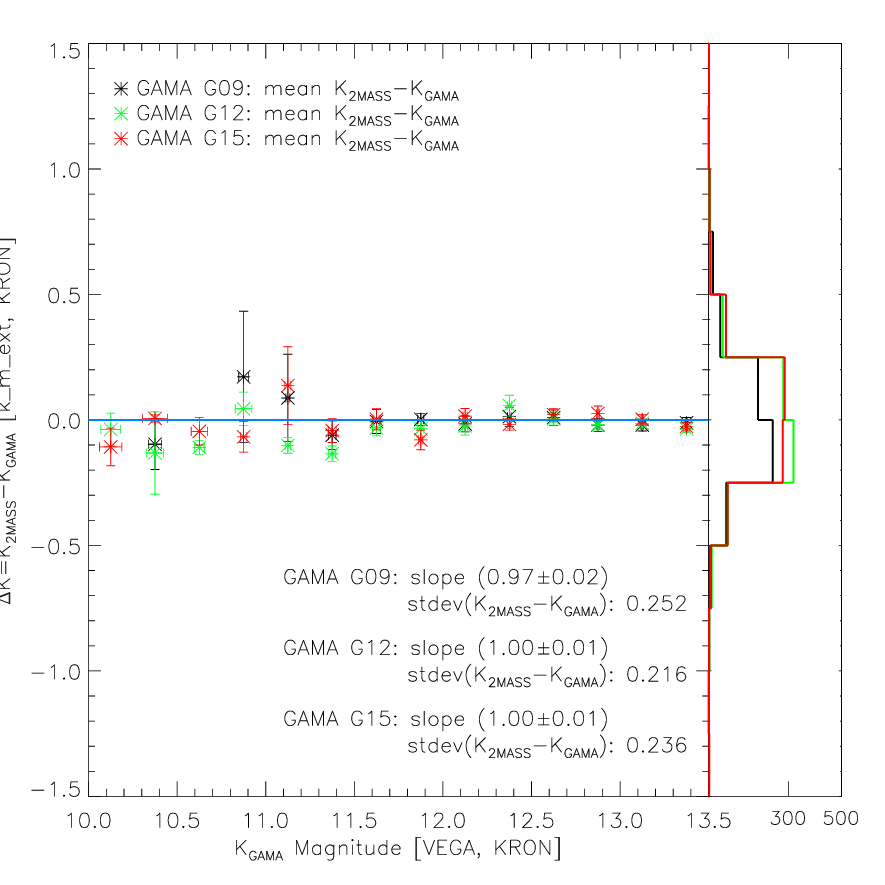}
 \caption{K band magnitude comparison between GAMA Kron and 2MASS k\_m\_ext
magnitudes over the GAMA regions. The derived slope using the `mpfitexy' routine
and the standard deviation of the residuals are stated.
}
 \label{fig:kk-gama-2mass}
\end{center}
\end{figure}

\begin{table}
 \begin{center}
 \begin{tabular}{ccc}

\hline\hline
Field & Number & $K_{2MASS}-K_{GAMA}$ \\
\hline\hline

G09 & 567 & ($-0.02\pm0.01$) \\

G12 & 750 & ($-0.03\pm0.01$) \\

G15 & 725 & ($-0.02\pm0.01$) \\

\hline\hline
\end{tabular}
\caption{A summary of the zeropoint corrections applied to the GAMA data to
calibrate onto the 2MASS photometric scale as derived using the `mpfiexy'
routine when assuming no scale error.
 }
\label{tb:gamaukidss_2mass_koffsets}
 \end{center}
\end{table}

\subsection{SDSS cmodel}

\begin{figure}
\begin{center}
 \includegraphics[height=7.25cm]{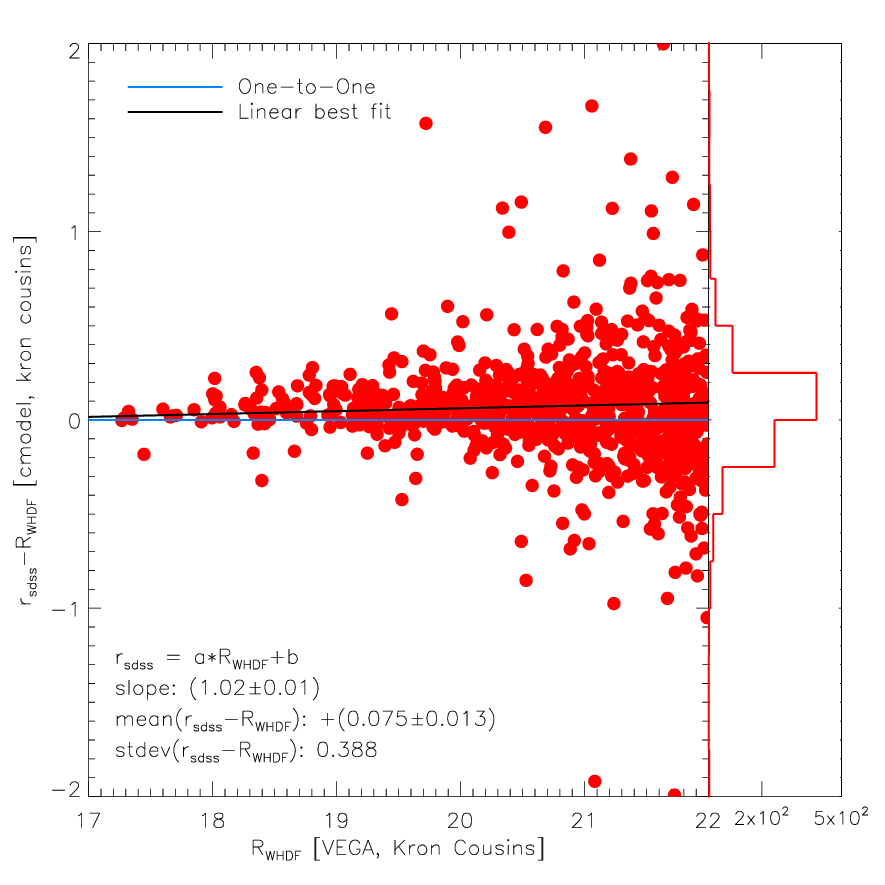}
 \caption{Magnitude comparison between WHDF Kron Cousins R and SDSS cmodel r
over the extended WHDF region. The derived slope
using the `mpfitexy' routine and both the mean and standard deviation of the
residuals are stated.
}
 \label{fig:rr-whdf-sdss}
\end{center}
\end{figure}

We now test the SDSS cmodel magnitude using Kron magnitudes from the extended
WHDF region Cousins R band data of \citet{metcalfe_2001,metcalfe_2006}. Although
some non-linearity is seen in Fig. \ref{fig:rr-whdf-sdss} this is due to
saturation of the WHDF bright magnitudes. In the range $17<r<22$ visually there
seem little evidence of a scale error and this is confirmed by an analysis using
`mpfitexy' where we find a slope of $r_{cmodel}$ = ($1.02\pm0.01$)$R_{WHDF}$. If
we then assume no scale error we find a simple zeropoint offset of
$r_{cmodel}$-$R_{WHDF}$=($0.07\pm0.01$)mag. However, for the SDSS r band count
in Fig. \ref{fig:optical-sdss-nm-euclidean} we have in fact assumed the larger
offset of $r_{cmodel}$-$R_{WHDF}$=0.12mag to ensure the counts at $r>21$ are in
agreement with the homogeneous model as might be expected at this depth.

\section{Incompleteness effects}
\label{append:magincompleteness}

\subsection{Photometric Incompleteness}

\subsubsection{2MASS}

2MASS is $\approx97.5\%$ complete to $K<13.57$ as described at
\path{http://www.ipac.caltech.edu/2mass/releases/allsky/doc/sec6_1k.html}.
Star-galaxy separation for $|b|>$20$\deg$ has been determined by eye to be $>$99\%
reliable to at least $K<12.8$ and only falling to 97\% by K$=$13.5 as outlined at
\path{http://www.ipac.caltech.edu/2mass/releases/allsky/doc/sec6_5b2.html}.

\subsubsection{SDSS}

The SDSS r band photometric catalogue is magnitude limited to $r<22.04$ and has
been validated by comparison to COMBO-17 as discussed at
\path{http://www.sdss3.org/dr9/imaging/other_info.php#completeness}. Any
significant incompleteness is only present at magnitudes r$>$21 which is far
fainter than the scales relevant for studying a local 300h$^{-1}$Mpc under-density.

Equally, SDSS have studied the validity of their star-galaxy separation relative
to COMBO-17 at
\path{http://www.sdss3.org/dr9/imaging/other_info.php#stargalaxy}. Significant
issues in classification arise at bright magnitudes $r<15$ and at faint
magnitudes r$>$20. Only the problem at the bright end is relevant for
interpreting the number counts at the Local Hole scales. However, the agreement
between the spectroscopically derived $\phi^*(z)$ models and the photometric
number counts suggests that star-galaxy separation is not biasing the bright end
results.

\subsection{Spectroscopic Incompleteness}

\begin{figure}
\begin{center}
 \includegraphics[height=7.25cm]{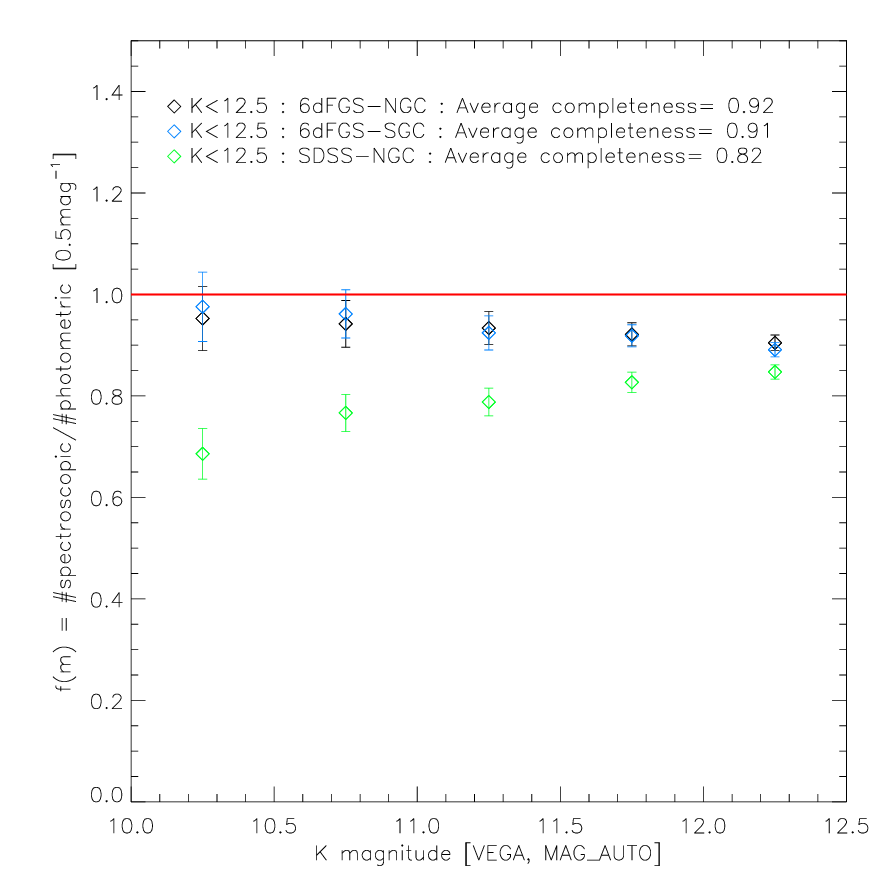}
 \caption{K band spectroscopic incompleteness as a function of magnitude as
derived from the ratio of spectroscopic (6dFGS \& SDSS) and photometric (2MASS) number counts with
$\delta m=0.1$. Poisson errors are shown.
}
 \label{fig:appen-magincompl-twomass}
\end{center}
\end{figure}

\begin{figure}
\begin{center}
 \includegraphics[height=7.25cm]{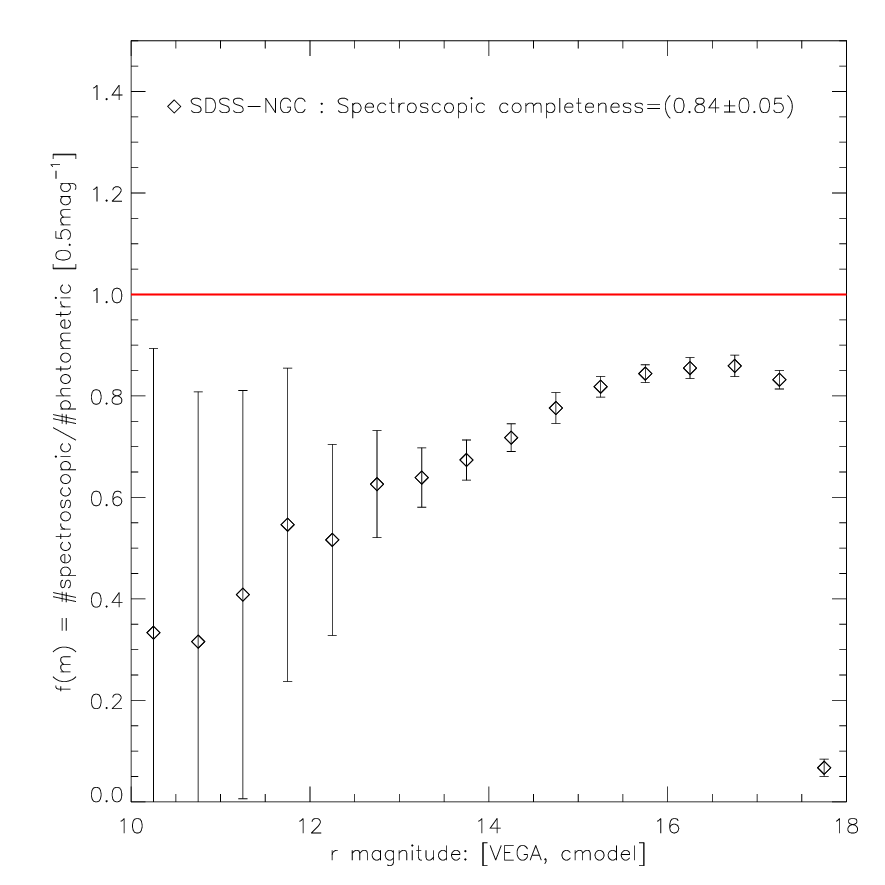}
 \caption{r band spectroscopic incompleteness as a function of magnitude as
derived from the ratio of spectroscopic (SDSS) and photometric (SDSS) number counts with
$\delta m=0.1$.  Poisson errors are shown.
}
 \label{fig:appen-magincompl-sdss}
\end{center}
\end{figure}

In Figs. \ref{fig:appen-magincompl-sdss} and \ref{fig:appen-magincompl-twomass}
we show respectively the spectroscopic incompleteness of the $K$ and $r$ samples
used in this paper. Also reported are the ratios of the total number of
spectroscopic to photometric galaxies for each sample. We can see that the
incompleteness increases for brighter galaxies, particularly in the case of the
r and K band SDSS-NGC samples. This is caused by the relative importance of
image artifacts and fibre-constraints for large/bright galaxies in SDSS
\citep{mcintosh_2006,bell_2003}.

We first correct the number of galaxies in the data $n(z)$ to the same total as
in the corresponding data $n(m)$ by multiplying the data $n(z)$ by the ratio of
the total number of photometric to spectroscopic galaxies in each sample. Next,
we account for the magnitude dependence of spectroscopic incompleteness in the
model $n(z)$ as shown in Figs. \ref{fig:appen-magincompl-twomass} and
\ref{fig:appen-magincompl-sdss}. We do this using introducing magnitude
dependent completeness factor $f(m)$ into the modelling procedure as in eq.
\ref{eq:redshiftdistimplement} by adjusting $\Phi(M)$ as follows,

\begin{align}
 \Phi(M) &\equiv \Phi(m-5\log d_{L}(z) -25 - K(z) - E(z)), \\
  &\rightarrow f(m) \Phi(m-5\log d_{l}(z) -25 - K(z) - E(z)), \nonumber
 \label{eq:append-magincompl}
\end{align}

\noindent while conserving galaxy numbers in the model $n(z)$. A similar
technique was then applied to correct $\overline{z}(m)$.

Finally, even at the low redshift end the change due to this procedure is less
than 1\% in the $n(z)$ for both the K and r limited spectroscopic datasets. It
is therefore irrelevant for interpreting the density profiles shown in Figs
\ref{fig:nz-ALL-phiratio}, \ref{fig:numbercount-norm-phiz} and
\ref{fig:optical-sdss-phiz}. However, the effect is somewhat more appreciable in
$\overline{z}(m)$, especially for the SDSS-NGC K sample where the completeness
correction can cause bins to vary by as much as 100 $kms^{-1}$. This is due to
the stronger variations in spectroscopic incompleteness for this sample.

\bsp
\label{lastpage}
\end{document}